\documentclass[runningheads]{llncs}

\usepackage{eccv}
\usepackage{eccvabbrv}
\usepackage{graphicx}
\usepackage{booktabs}

\usepackage[accsupp]{axessibility}

\usepackage{xspace}
\usepackage{graphicx}
\usepackage{colortbl}
\usepackage{xcolor}
\usepackage{array}
\usepackage{booktabs}
\usepackage{multirow}
\usepackage{caption}
\usepackage{subcaption}
\usepackage{amssymb}
\usepackage{hhline}
\usepackage{nicefrac}       
\usepackage{microtype}      
\usepackage{ragged2e}
\usepackage{url}

\usepackage{mdframed}
\definecolor{blockquote-bg}{gray}{0.95}   
\definecolor{blockquote-bar}{gray}{0.70}  

\newmdenv[
  backgroundcolor=blockquote-bg,
  linecolor=blockquote-bar,
  leftline=true,
  rightline=false,
  topline=false,
  bottomline=false,
  linewidth=3pt,
  innerleftmargin=10pt,
  innerrightmargin=10pt,
  innertopmargin=6pt,
  innerbottommargin=6pt,
  roundcorner=2pt,        
  skipabove=\baselineskip,
  skipbelow=\baselineskip
]{blockquote}

\definecolor{ctcolor}{HTML}{D4E4FA}
\definecolor{dermcolor}{HTML}{D9F3FD}
\definecolor{endocolor}{HTML}{D5F6F2}
\definecolor{funduscolor}{HTML}{D9F5D6}
\definecolor{mrcolor}{HTML}{EEF7C7}
\definecolor{pathologycolor}{HTML}{FAF1D1}
\definecolor{uscolor}{HTML}{FEDCB4}
\definecolor{xraycolor}{HTML}{FBC8BC}
\definecolor{avgcolor}{HTML}{F5F5F5}

\newcommand{\methodname}{MedITok\xspace}

\definecolor{longzilipurple}{RGB}{102, 51, 153}

\newcommand{\mat}[1]{\boldsymbol{#1}} 
\newlength\savewidth

\usepackage{appendix}

\usepackage{diagbox}
\usepackage{etoc}
\etocdepthtag.toc{mtchapter}
\etocsettagdepth{mtchapter}{subsection}
\etocsettagdepth{mtappendix}{none}

\usepackage{xcolor}
\usepackage{tikz}
\usepackage{wrapfig}
\usetikzlibrary{patterns}

\NewDocumentCommand{\flexsquare}{ O{red} O{true} }{%
  \begingroup
    \tikz[baseline=(sq.south)]{%
      \ifthenelse{\equal{#2}{true}}{%
        \node[
          minimum size=0.8em,
          inner sep=0pt,
          draw=black,
          pattern=north east lines,
          pattern color=#1
        ] (sq) {};
      }{%
        \node[
          minimum size=0.8em,
          inner sep=0pt,
          draw=black,
          fill=#1
        ] (sq) {};
      }%
    }%
  \endgroup
}

\newcommand{\sortcites}[1]{%
  \begingroup
  \def\NAT@sort@cites##1{\NAT@sort{##1}}
  \NAT@sort@cites{\cite{#1}}
  \endgroup
}
\usepackage[breaklinks,colorlinks,allcolors=eccvblue]{hyperref}
\usepackage{orcidlink}

\begin{document}

\title{Unified Medical Image Tokenizer for Autoregressive Synthesis and Understanding}

\titlerunning{Unified Medical Image Tokenizer for Autoregressive Synthesis and Understanding}

\author{
Chenglong Ma$^{\dagger}$\textsuperscript{1,2} \and
Yuanfeng Ji$^{\dagger}$\textsuperscript{3} \and
Jin Ye\textsuperscript{4} \and
Zilong Li\textsuperscript{1} \and  
Chenhui Wang\textsuperscript{1} \and \\ 
Junzhi Ning\textsuperscript{4}  \and
Wei Li\textsuperscript{4} \and  
Lihao Liu\textsuperscript{4} \and  
Qiushan Guo\textsuperscript{5} \and  
Tianbin Li\textsuperscript{4}  \and \\
Junjun He$^{*}$\textsuperscript{2,4} \and 
Hongming Shan$^{*}$\textsuperscript{1} \\
\vspace{5pt}
\textsuperscript{1}Fudan University \quad
\textsuperscript{2}Shanghai Innovation Institute \quad
\textsuperscript{3}Stanford University \\
\textsuperscript{4}Shanghai AI Laboratory \quad
\textsuperscript{5}ByteDance Seed
}

\authorrunning{C.~Ma et al.}

\institute{}

\begingroup
\renewcommand{\thefootnote}{}
\footnotetext{$^\dagger$Equal contribution. $^*$Co-correspondence.}
\endgroup

\maketitle

\begin{abstract}
Autoregressive modeling has driven major advances in multimodal AI, yet its application to medical imaging remains constrained by the absence of a \emph{unified image tokenizer} that simultaneously preserves fine-grained anatomical structures and rich clinical semantics across heterogeneous modalities.
Existing approaches jointly optimize image reconstruction and textual semantic objectives, relying on large-scale image-caption pairs and are prone to gradient interference. This is ill-suited for the medical domain where paired data are scarce and abundant unpaired images remain unexploited.
This work identifies these issues in building unified medical image tokenizers, and introduces a principled two-stage training framework using \emph{visual representation as a bridge} to address them. The propose visual representation alignment stage enables the utilization of large-scale unpaired medical images to ensure reconstruction fidelity and establish foundational semantics, alleviating the interference and better preparing for the second stage where fine-grained textual semantics are injected using image-text pairs.
The resulting tokenizer, \methodname, is trained on over 33 million medical images spanning 9 modalities and 2 million image-text pairs. 
\methodname achieves state-of-the-art performance on  30+ benchmarks spanning 9 imaging modalities and 4 task families. It further enables autoregressive modeling for diagnostic and generative applications, serving as a scalable component for future multimodal models with unified synthesis and understanding capabilities in the medical domain. Project page: \url{https://github.com/Masaaki-75/meditok}

\end{abstract}

\section{Introduction}\label{sec:intro}

The rapid evolution of advanced autoregressive (AR) models, such as GPT-4o~\cite{openai2025nativeimage}, has revolutionized multimodal learning. These models excel at generating and understanding text, image, and audio data via unified processing of token-based representations. 
In medical imaging, AR models begin to demonstrate similar promise, powering report generation~\cite{tanno2025flamingocxr}, tumor segmentation~\cite{chen2025autoregressive}, counterfactual synthesis~\cite{ma2025towards}, and diagnostic visual question answering (VQA)~\cite{li2023llavamed}. By translating complex biomedical image patterns into token sequences, these models can synthesize realistic images and interpret clinical cues (\eg, ground-glass opacities on chest computed tomography, 
pleural effusions on X-rays) in the images, with the potential to streamline workflows and improve patient outcomes.

A critical ingredient in building a powerful AR model is the \emph{image tokenizer}, which translates an input image to a sequence of discrete tokens suitable for AR modeling. Existing approaches can be divided into two categories. (1) Synthesis‐oriented tokenizers optimized for pixel-level reconstruction\footnote{In this paper, ``reconstruction'' refers to autoencoding reconstruction: decoding an input image from its latent representation.}, \eg, VQGAN~\cite{esser2021taming}. These tokenizers precisely capture low-level structure in the image that is vital to image compression~\cite{varma2025medvae,wang2024devil} and generation~\cite{zhu2024scaling,sun2024llamagen,yu2024tokenizeriskey,yao2025reconstruction}. However, they do not explicitly encode discriminative features and are therefore not suitable for interpreting the concepts and objects embedded in the image. 
(2) Understanding‐driven tokenizers trained with discriminative objectives, \eg, CLIP~\cite{radford2021clip}. These tokenizers excel at capturing high-level textual semantics, making them indispensable for visual comprehension, but they fail to accurately retain spatial structures and textures in the image. 

\noindent\textbf{Motivation.}\quad
Image tokens that embed only one side of this structure-semantic spectrum will offload the representation learning burden onto downstream AR models, which often incurs heavy pre-training costs and can still leave either side under-utilized~\cite{wang2024emu3,chen2025janus}. 
These limitations are especially acute in the medical domain, where clinical tasks typically demand both precise visual structures and clinical semantics. 
However, current medical image tokenizers tend to specialize in one single aspect~\cite{luo2023pumit,zhang2023biomedclip,zhang2022contrastive,huang2021gloria}, which lacks a unified, information-rich token space and thereby limits the potential of downstream medical AR models for accurate and data-efficient diagnosis.

Our goal is to democratize a foundation unified tokenizer for medical images. 
Nonetheless, training a unified tokenizer for medical images poses unique challenges: (1) a na\"ive joint optimization of image reconstruction and textual semantic objectives may introduce mutual interference and suboptimal performance~\cite{wu2025vilau,qu2024tokenflow};  and (2) paired image-caption data for training is much more scarce in the medical domain, compared to the abundant unlabeled images.

To addresses these issues, we propose a novel two-stage training framework. Instead of directly coupling the visual reconstruction and textual semantic, it involves a \emph{visual representation alignment stage as a bridge} to first establish basic semantic awareness with reconstruction capabilities, followed by the textual semantic alignment stage for learning finer-grained semantic information. 
This framework leads to our model: \methodname, the first unified medical image tokenizer. \methodname encodes both low-level structural information, supporting faithful image reconstruction and realistic image synthesis, and high-level semantics, enabling multimodal medical image comprehension, serving as a foundation for diverse community use. 

Specifically, the first training stage trains \methodname on pure medical images, optimizing for reconstruction fidelity with a visual semantic constraint on the latent space. Then, the textual semantic alignment stage tunes \methodname on image-caption pairs, enhancing semantic richness by aligning image tokens to fine-grained textual embeddings of detailed captions. 
This approach allows \methodname to \emph{effectively encode structural and semantic information} while \emph{exploiting both unpaired medical images and image-text pairs at scale}, making a unified latent space to develop powerful AR models for diverse tasks. 
To achieve this, we meticulously collect a large-scale dataset comprising over 30 million medical images and 2 million image-caption pairs from multiple public sources, with broad coverage of imaging modalities, anatomies, and pathologies. This collection ensures that \methodname learns robust representations for medical images across diverse clinical contexts. 

\noindent\textbf{Contributions.}\quad
\textbf{(1)} We propose a novel training framework for developing a unified image tokenizer, which effectively scales up with medical image data and progressively builds a unified latent space. 
\textbf{(2)} We introduce \methodname, the first Medical Image Tokenizer that unifies the encoding of low-level structural details and high-level clinical semantics. 
\textbf{(3)} Extensive experimental results on over 30 datasets, spanning 9 imaging modalities, across 4 different task families, showcase \methodname's state-of-the-art performance over existing approaches and broad applicability to diverse medical tasks. 
\textbf{(4)} Model and code will be open-source. Data accesses are provided respecting all original licenses.

\section{Related Work}\label{sec:related}

\noindent\textbf{AR Models in Medical Image Tasks.}\quad
AR models have shown impressive scalability and generalizability in general vision-language processing. In medical domain, these models have been extended to a variety of tasks: image captioning and VQA for interpreting scans and assist diagnosis~\cite{li2023llavamed,moor2023medflamingo,chen2024huatuogpt}, lesion segmentation model across different imaging modalities~\cite{chen2025autoregressive}, medical image synthesis for counterfactual analysis~\cite{ma2025towards} and modality transfer~\cite{ren2024medical}, \etc 
More recently, HealthGPT~\cite{lin2025healthgpt}, empowered by the VQGAN as the image tokenizer, further unifies both medical image synthesis and comprehension capabilities within an AR framework for broader applications. However, these methods typically general-domain tokenizers pre-trained on natural images, which encode insufficient clinical knowledge and capture either low-level structural detail or high-level clinical concepts, rarely both, whereas clinical tasks usually demand joint representation. To this end, we introduce \methodname, a unified tokenizer tailored for medical images to support a wide range of tasks and empower advanced AR models in the medical field. 


\noindent\textbf{Unified Image Tokenizers.}\quad
Image tokenizers encode images into token sequences suitable for AR modeling. Recent works~\cite{wu2025vilau,ma2025unitok,qu2024tokenflow} seek to unify the encoding of both low-level details and high-level semantics into one single tokenizer, enhancing the multimodal generation and comprehension capabilities of downstream AR models. TokenFlow~\cite{qu2024tokenflow} proposes an intuitive dual-codebook design that explicitly decouples semantic and pixel-level cues. UniTok~\cite{ma2025unitok} instead shows that simply scaling codebook capacity improves the unified encoding capability. 
Our approach proposes to use pretrained representational image encoder as a bridge to ease the unified encoding of structural and semantic information, while effectively utilizing the abundant unlabeled images.
This offers a unified latent space to power the next generation of medical multimodal models. Please refer to Appendix~\ref{sec:supp:discussion-related} 
for more discussion.

\begin{figure*}[t]
    \centering
    \includegraphics[width=\linewidth]{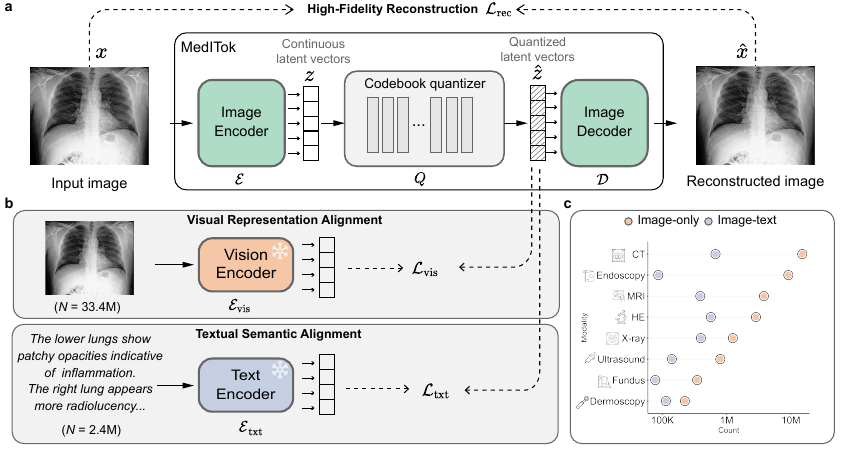}
    \caption{Overview of the proposed training framework. (a) Architecture of \methodname. (b) Two-stage training: visual representation alignment with pretrained visual semantics, followed by textual semantic alignment using clinical captions. (c) Statistics across modalities for our training data.} 
    \label{fig:method}
\end{figure*}

\section{Methodology}\label{sec:method}

By encoding both low-level structural details and high-level clinical semantics, \methodname directly supports medical image reconstruction and classification tasks, and can be further integrated into AR models for more complex tasks, \eg, medical image synthesis and understanding, \etc 
Below, we start with a preliminary on the image tokenization (Sec.~\ref{sec:method-pre}) and provide detailed description of our model and training framework (Sec.~\ref{sec:method-model}) and dataset curation process (Sec.~\ref{sec:method-data}).

\subsection{Preliminary}\label{sec:method-pre}

The drive to apply powerful AR models from natural language processing to visual data has spurred the development of image tokenization techniques, converting images into sequences of tokens. Among these, Vector Quantization (VQ)-based approaches~\cite{van2017vqvae,esser2021taming} are foundational.

In a typical VQ-based image tokenizer, an image $\mat{x}$ is first mapped by an encoder $\mathcal{E}$ to a spatial grid of latent vectors $\mat{z} \in \mathbb{R}^{h \times w \times d}$. Each vector in this grid is then quantized by assigning it to the closest entry in a learned, finite codebook $\mathcal{C} = \{\mat{c}_k\}_{k=1}^{K}$, where $\mat{c}_k \in \mathbb{R}^{d}$ represents an image token and $K$ is the codebook size. The quantized grid of latent vectors, $\mat{z}_\mathrm{q} \in \mathbb{R}^{h \times w \times d}$, effectively represent the image as a compressed sequence of image tokens. A decoder $\mathcal{D}$ is then trained to reconstruct the image from these representations, producing $\hat{\mat{x}}=\mathcal{D}(\mat{z}_\mathrm{q})$.
During training, the encoder $\mathcal{E}$, decoder $\mathcal{D}$, and the codebook $\mathcal{C}$ are jointly optimized. It typically involves a composite loss function designed to ensure both accurate reconstruction and effective codebook learning~\cite{esser2021taming}, defined as:
\begin{align}
    \mathcal{L}_\mathrm{rec}(\hat{\mat{x}}, \mat{x}, \mat{z}_\mathrm{q}, \mat{z}) &= \mathcal{L}_\mathrm{img}(\hat{\mat{x}},  \mat{x}) 
    + \lambda_\mathrm{vq} \mathcal{L}_\mathrm{vq}(\mat{z}_\mathrm{q}, \mat{z}),\\
    \mathcal{L}_\mathrm{vq}(\mat{z}_\mathrm{q}, \mat{z}) & = 
    \Vert \mat{z}_\mathrm{q} - \mathrm{sg}[\mat{z}] \Vert_2^2 
    + \beta\Vert \mathrm{sg}[\mat{z}_\mathrm{q}] - \mat{z} \Vert_2^2,
    \label{eq:vq-loss}
\end{align}
where $\mathcal{L}_\mathrm{img}$ is the image fidelity loss consisting of a mean square error loss, a perceptual loss~\cite{johnson2016perceptual}, and an adversarial loss, encouraging high-fidelity reconstructions. The vector quantization loss~\cite{van2017vqvae} $\mathcal{L}_\mathrm{vq}$ ensures the encoder outputs $\mat{z}$ to commit to their nearest codebook vectors $\mat{z}_\mathrm{q}$, with a factor $\beta$ and the stop-gradient operation $\mathrm{sg[\cdot]}$. 
Our work builds upon these foundational principles of VQ-based tokenization but introduces a novel training framework tailored to unified medical image tokenization.

\subsection{\methodname Training Framework}\label{sec:method-model}

A unified image tokenizer must reconcile two objectives that naturally compete: preserving low-level spatial detail for image synthesis, and learning a high-level semantic token space for image understanding. Previous works~\cite{wu2025vilau,ma2025unitok} combine image reconstruction and textual representation learning objectives in one go. Such training can lead to representation collapse or suboptimal trade‑offs~\cite{qu2024tokenflow}. Moreover, they typically rely on large-scale image-text pairs while overlooking the abundance of unpaired images. We propose a novel two-stage training framework to train our unified image tokenizer \methodname, unlocking the potential of unlabeled images in the medical domain and progressively transitioning from reconstruction-focused learning to unified token learning.

As depicted in Fig.~\ref{fig:method}, \methodname is comprised of an image encoder $\mathcal{E}$, a quantizer $Q$, and a decoder $\mathcal{D}$. 
Our framework begins with a \emph{visual representation alignment} stage, which cold‑starts the training of the image encoder $\mathcal{E}$ and a decoder $\mathcal{D}$ using a vast corpus of unpaired medical images. The primary focus is on capturing low-level structural information, guided by only a light semantic constraint from a pretrained vision encoder $\mathcal{E}_\mathrm{vis}$. Subsequently, in the second stage termed \emph{textual semantic alignment}, $\mathcal{E}$ is refined using image-text pairs. Here, the emphasis shifts towards enhancing the semantic richness of the learned tokens by aligning them with clinical captions processed by a pretrained text encoder $\mathcal{E}_\mathrm{txt}$. This approach not only alleviates the conflicts between reconstruction and semantic learning objectives but also allows us to effectively leverage large-scale unpaired images alongside paired image-text data for unified tokenizer training. 

\noindent\textbf{Visual Representation Alignment (S1).}\quad
Given an input image $\mat{x}$, the encoder $\mathcal{E}$ produces continuous latent vectors $\mat{z}$, which are then quantized by the quantizer $Q$ to yield discrete latent vectors $\mat{z}_\mathrm{q} = Q(\mat{z})$. 
The decoder $\mathcal{D}$ then learns to reconstruct the image $\hat{\mat{x}} = \mathcal{D}(\mat{z}_\mathrm{q})$. The pretrained vision encoder $\mathcal{E}_\mathrm{vis}$ encodes the image $\mat{x}$ into a semantic representation, which is then projected into the space of $\mat{z}_\mathrm{q}$ via a linear layer $f_\mathrm{vis}$ to provide semantic supervision for learning $\mat{z}_\mathrm{q}$. 
We use a composite loss function for training, defined as:
\begin{align}
    \mathcal{L}_\mathrm{S1} \!=\! \mathcal{L}_\mathrm{rec}(\hat{\mat{x}}, \mat{x}, \mat{z}_\mathrm{q}, \mat{z}) 
    \!+\! \lambda_\mathrm{vis} \mathcal{L}_\mathrm{vis}(\mat{z}_\mathrm{q}, f_\mathrm{vis}
    (\mathcal{E}_\mathrm{vis}(\mat{x}))),
    \label{eq:stage1-loss}
\end{align}
where $\mathcal{L}_\mathrm{vis}$ is a contrastive loss that imposes light semantic constraint on the latent space, with the factor $\lambda_\mathrm{vis}$ set to 0.1. 
By ensuring reconstruction while gently guiding the latent space with pre-trained visual semantics, this stage helps \methodname develop a robust encoding of visual structure, preparing it for fine-grained semantic alignment in the subsequent stage.

\noindent\textbf{Textual Semantic Alignment (S2).}\quad
This stage focuses on enhancing the semantic richness of the learned image tokens and aligning them with fine-grained textual representations extracted from detailed clinical captions. The training in this stage is driven by the following loss function: 
\begin{align}
    \mathcal{L}_\mathrm{S2} \!=\! \mathcal{L}_\mathrm{rec}(\hat{\mat{x}}, \mat{x}, \mat{z}_\mathrm{q}, \mat{z}) 
    \!+\! \lambda_\mathrm{txt} \mathcal{L}_\mathrm{txt}(\mat{z}_\mathrm{q}, 
    f_\mathrm{txt}
    (\mathcal{E}_\mathrm{txt}(\mat{t}))),
    \label{eq:stage2-loss}
\end{align}
where $\mat{t}$ denotes the caption of the image $\mat{x}$, and $f_\mathrm{txt}$ is another linear layer. $\mathcal{L}_\mathrm{txt}$ is the contrastive loss, with a balancing factor $\lambda_\mathrm{txt}$ set to 1.
This stage further integrates the structural and semantic representation learning, empowering \methodname for a wide range of downstream medical applications requiring nuanced understanding. 

\subsection{Dataset Curation}\label{sec:method-data}

The development of \methodname necessitates a comprehensive and diverse dataset. To this end, we aggregate medical images and image-text pairs from existing publicly available sources. 
For example, image-text pairs are collected from BIOMEDICA~\cite{lozano2025biomedica}, MedICaT~\cite{subramanian2020medicat}, MIMIC-CXR~\cite{johnson2019mimic}, ROCOv2~\cite{ruckert2024rocov2}, PMC-OA~\cite{lin2023pmcoa}, MM-Retinal~\cite{wu2024mmretinal}, and GMAI-MM-Caption-1.7M~\cite{li2024gmai} datasets. These datasets extract and filter image-text pairs from raw clinical imaging, literatures, clinical reports, \etc, ensuring data diversity. 

To further ensure that the training data are of sufficient quality for learning meaningful representations, after collecting these data, we perform a rigorous data quality control with a combination of automated and manual filtering to exclude images of low quality or limited clinical relevance. Specifically, an image is excluded if, after proxy RGB conversion, it meets any of the following criteria: 
(1) low pixel intensity range below 50; 
(2) insufficient resolution, where the smallest dimension is under 128 pixels; 
(3) low information content, indicated by a standard deviation of pixel values below 10; 
(4) limited palette, with three or fewer unique pixel values; and 
(5) unrelated content, such as tables, plots, or non-clinical images extracted from publications. 
For text data, we only retain texts pertinent to clinical imaging, determined by the tags and metadata within each dataset. 

These checks efficiently remove noisy and uninformative samples and ensures higher quality input for our training framework, resulting in a massive corpus of 33,428,922 medical images for the visual representation alignment stage, and 2,422,827 high-quality medical image-text pairs for the textual semantic alignment stage. This collection encompasses over eight imaging modalities, including computed tomography (CT), dermoscopy, endoscopy, fundus photography, magnetic resonance imaging (MRI), pathology, ultrasound, and X-ray, spanning a wide spectrum of anatomical regions and pathological findings. We leave detailed sources and statistics in our Appendix~\ref{sec:supp:training-dataset}.

\section{Experiments}\label{sec:exp}

We present comprehensive experiments to evaluate the proposed \methodname across four different task families, including low- and high-level encoding, as well as medical image synthesis and understanding. Since each task is evaluated using specialized datasets and metrics appropriate to its goals, we introduce them within each corresponding subsection. 
We conduct deduplication and manual cross-checking to ensure no overlap between data used for evaluating and training \methodname. 
Please see Appendix~\ref{sec:supp:benchmarking-datasets} and~\ref{sec:supp:setup} 
for more details on dataset statistics, tasks, and implementation.

\subsection{Experimental Setup}

\begin{table*}[!t]
\centering
\caption{Medical image reconstruction across different imaging modalities using different models. The best results are highlighted in \textbf{bold} and the second best results are \underline{underlined}. SSIM values are presented as percentages. $f_\mathrm{d}$ denotes the downsampling factor. ``$\downarrow$'': The lower the better. ``$\uparrow$'': The higher the better. 
}
\small
\resizebox{1.0\textwidth}{!}{%
\begin{tabular}{llccccccccc|cc}
\toprule
Metrics & Models & $f_\mathrm{d}$ & CT & Derm. & Endo. & Fundus. & MRI & Path. & US & X-ray & Avg. & Avg. rank \\
\midrule
\multirow{6}{*}{rFID ($\downarrow$)} 
& VQGAN & 8 & 15.97 & 33.57 & 27.33 & 27.22 & 21.33 & 67.68 & 29.48 & 18.66 & 30.16 & 4.9 \\
& Emu3-VQ & 8 & 11.83 & 27.91 & 20.83 & 16.27 & 13.52 & 69.89 & 25.43 & 11.99 & 24.71 & 3.4 \\
& VAR-VQ & 16 & 14.69 & 30.27 & 19.74 & 21.69 & 13.99 & 70.06 & 26.09 & 12.18 & 26.09 & 4.1 \\
& TokenFlow & 16 & 24.78 & 44.28 & 47.42 & 34.93 & 26.81 & 98.22 & 51.77 & 24.51 & 44.09 & 7.0 \\
& UniTok & 16 & \underline{9.27} & \underline{23.15} & \underline{13.64} & \underline{16.22} & \underline{9.30} & \underline{47.77} & \underline{20.93} & \underline{8.61} & \underline{18.61} & \underline{2.0} \\
\cmidrule(lr){2-13}
& PUMIT &16 & 32.67 & 53.46 & 56.22 & 27.51 & 25.43 & 142.98 & 37.04 & 23.78 & 49.88 & 7.1 \\
& MedVAE & 8 & 20.17 & 140.39 & 114.00 & 117.39 & 23.34 & 123.20 & 30.60 & 11.54 & 73.64 & 6.5 \\
\rowcolor{gray!20} & \methodname & 16 & \textbf{7.88} & \textbf{22.27} & \textbf{10.66} & \textbf{14.39} & \textbf{6.32} & \textbf{46.54} & \textbf{17.64} & \textbf{6.55} & \textbf{16.53} & \textbf{1.0} \\
\midrule

\multirow{6}{*}{PSNR ($\uparrow$)} 
& VQGAN & 8 & 31.13 & 29.28 & 25.60 & 35.40 & 29.54 & 20.42  & 24.79 & 31.68 & 28.48 & 6.3 \\
& Emu3-VQ & 8 & 36.11 & \underline{31.68} & 28.96 & \textbf{39.64} & \underline{34.32} & 22.08  & 27.57 & \underline{35.81} & \textbf{32.02} & \underline{2.6} \\
& VAR-VQ & 16 & 31.32 & 29.26 & 25.75 & 35.73 & 29.83 & 20.86 & 25.22 & 31.10 & 28.63 & 5.8 \\
& TokenFlow & 16 & 28.64 & 27.23 & 23.72 & 33.45 & 27.68 & 19.33  & 23.26 & 28.71 & 26.50 & 7.8 \\
& UniTok & 16 & 33.60 & 30.97 & 27.55 & 37.21 & 31.50 & 22.18  & 26.97 & 32.97 & 30.34 & 4.3\\
\cmidrule(lr){2-13}
& PUMIT &16 & 33.64 & 30.23 & \underline{29.08} & 37.33 & 33.13 & \underline{23.09} & 28.31 & 33.89 & 31.09 & 3.1 \\
& MedVAE & 8 & \textbf{36.46} & 20.67 & 25.04 & 15.31 & \textbf{34.42} & 19.58  & \underline{28.29} & \textbf{36.23} & 27.01 &4.5  \\
\rowcolor{gray!20} & \methodname & 16 & \underline{36.32} & \textbf{31.69} & \textbf{29.19} & \underline{37.72} & 33.55 & \textbf{23.54}  & \textbf{28.49} & 34.42 & \underline{31.74} & \textbf{1.8} \\
\midrule 

\multirow{6}{*}{SSIM ($\uparrow$)} 
& VQGAN & 8 & 88.51 & 75.28 & 76.84 & 92.32 & 84.39 & 48.42  & 68.18 & 91.14 & 78.14 & 6.8 \\
& Emu3-VQ & 8 & 92.79 & 79.34 & 84.71 & 94.33  & 95.72 & 54.70 & 75.14 & \textbf{95.54} & 83.78 &3.5 \\
& VAR-VQ & 16 & 89.51 & 76.69 & 79.21 & 93.08 & 93.68 & 47.40 & 69.99 & 90.79 & 80.04 & 6.0\\
& TokenFlow & 16 & 82.43 & 67.19 & 69.47 & 89.60 & 90.22 & 33.09 & 56.56 & 84.50 & 71.63 &7.8 \\
& UniTok & 16 & 92.42 & \underline{81.00} & 84.47 & 94.45 & 95.47 & 56.42 & 76.40 & 92.74 & 84.17 &3.9 \\
\cmidrule(lr){2-13}
& PUMIT &16 & 92.10 & \underline{85.41} & \underline{87.81} & \underline{94.60} & \underline{96.59} & \underline{63.81} & \underline{81.46} & 94.52 & \underline{87.04} & \underline{2.6} \\
& MedVAE & 8 & \underline{92.86} & 75.32 & 81.52 & 69.46 & 95.92 & 53.10 & 77.45 & 94.77 & 80.10 &4.4 \\
\rowcolor{gray!20} & \methodname & 16 & \textbf{93.73} & \textbf{85.47} & \textbf{88.99} & \textbf{95.27} & \textbf{97.22} & \textbf{65.99} & \textbf{83.93} & \underline{95.39} & \textbf{88.25} & \textbf{1.1} \\

\bottomrule
\label{tab:main-recon}
\end{tabular}
}
\end{table*}

\noindent\textbf{Implementation Detail.}\quad
We build \methodname with a hybrid ViT architecture~\cite{chen2024vitamin} using PyTorch~\cite{pytorch}, and implement the quantizer with 32,768 eight-dimensional codebook vectors. We train \methodname using AdamW~\cite{loshchilov2018adamw} optimizer for 200k steps in the first stage and 10k steps in the second stage, with a global batch size of 512. 
Image resolution is of 256 $\times$ 256. The encoder of \methodname is initialize with weights from UniTok for efficient training. We choose BiomedCLIP~\cite{zhang2023biomedclip} as the pretrained semantic vision and text encoders for alignment in our training framework, which is frozen throughout the training. 
\noindent\textbf{Competing Tokenizers.}\quad
We compare \methodname with powerful image tokenizers from both general and medical domains, including VQGAN~\cite{esser2021taming}, Emu3-VQ~\cite{wang2024emu3}, VAR-VQ~\cite{tian2024var}, TokenFlow~\cite{qu2024tokenflow}, UniTok~\cite{ma2025unitok}, PUMIT~\cite{luo2023pumit}, and MedVAE~\cite{varma2025medvae}. VQGAN, Emu3-VQ, and VAR-VQ are pure VQ-based tokenizers trained on natural images without semantic alignments, yet have shown promising results in building medical multimodal models~\cite{lin2025healthgpt,ma2025towards}. TokenFlow and UniTok are two state-of-the-art unified image tokenizers in the general domains. PUMIT and MedVAE are medical image tokenizers that excel at visual fidelity and detail preservation.

\begin{figure}[th]
    \centering
    \includegraphics[width=\linewidth]{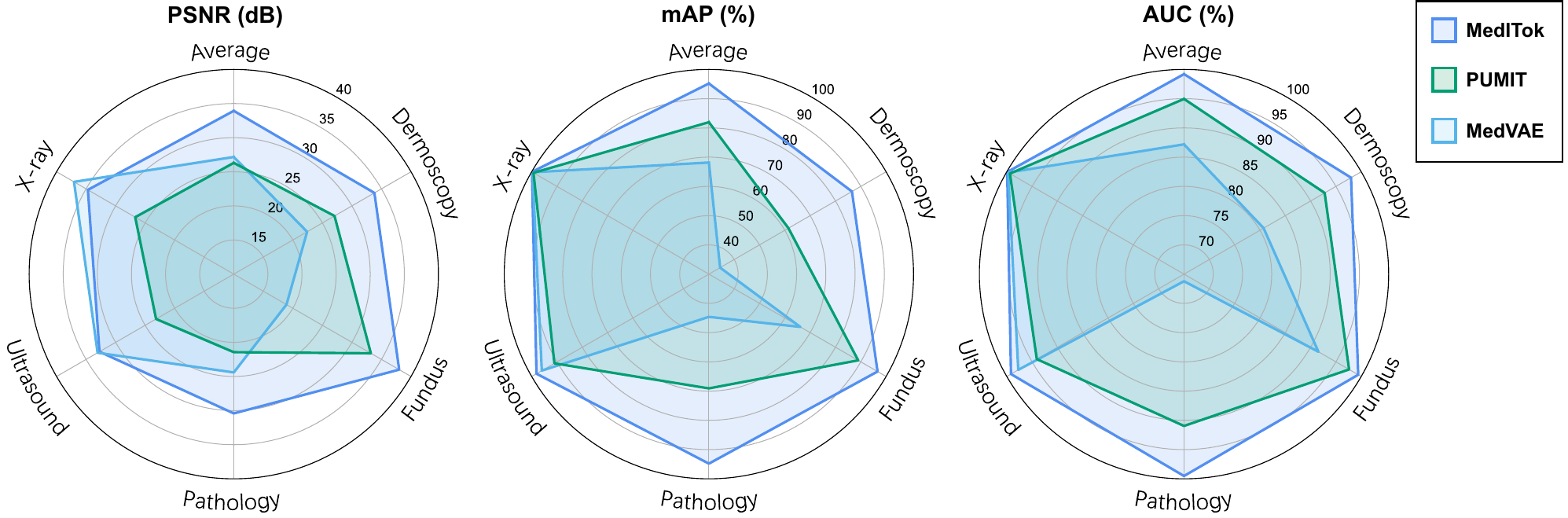}
    \caption{Diagnostic information preservation comparison.}
    \label{fig:reconcls-radar}
\end{figure}

\subsection{Low-Level Encoding Evaluation}\label{sec:exp-recon}
We evaluate how much low-level detail is encoded in the latent space of each tokenizer via image reconstruction task from two perspectives: the image fidelity and the diagnostic information preservation.

\noindent\textbf{Image Fidelity.}\quad 
We directly employ reconstruction Fr\'echet inception distance (rFID)~\cite{heusel2017fid}, peak signal-to-noise ratio (PSNR), and structural similarity index measure (SSIM)~\cite{wang2004ssim} to evaluate the image reconstruction performance. Notably, Woodland \etal~\cite{woodland2024fidmed} have shown that ImageNet-pretrained feature extractors are more consistent and aligned with human medical expert judgment than their counterparts pretrained on medical images, and we follow their work to implement rFID. 
In this part, we collect images from 23 publicly available datasets~\cite{mccollough2017aapmldct,landman2015btcv,heimann2009silver07,kawahara2018derm7pt,giotis2015mednode,ali2022monkeypox,kiranyaz2020hmcqu,cartucho2024surgt,ali2020endocv20edd,decenciere2014messidor,ovreiu2021acrima,fraz2012chasedb,hoover2000stare,graham2019consep,da2022digestpath,nir2018gleason,bao2025hie,pati2020ivygapradiomics,pedraza2015ddti,al2020busi,lian2021chestxdet,halabi2019rsnabone,tabik2020covidgr}, totaling 35,736 images covering 8 imaging modalities for evaluation. 

Quantitative results are shown in \cref{tab:main-recon}. 
MedVAE struggles on the modalities with colored imaging (\eg, fundus photography) as it is trained only on grayscale radiological images~\cite{varma2025medvae}. Notably, despite with a large downsampling factor of 16$\times$, \methodname delivers the best overall fidelity across 8 modalities, outperforming other tokenizers including those with only 8$\times$ downsampling. This highlights the efficiency of \methodname in balancing compression with reconstruction fidelity.

\noindent\textbf{Diagnostic Information Preservation.}\quad 
Beyond conventional metrics like PSNR and SSIM, we further evaluate the diagnostic information preservation in the images reconstructed by different tokenizers via a proxy classification task. 
Specifically, for a medical image classification dataset, we first run an image tokenizer to reconstruct images in its test split. Then, we train a ResNet using real images in the training split, and then compute the mean average precision (mAP) and area under the ROC curve (AUC) by comparing ResNet's classification of real test images and that of the test images reconstructed by the tokenizer. Higher mAP and AUC values indicate more preserved diagnostic information. 

To this end, we use five datasets in different imaging modalities, including 
ISIC2018~\cite{tschandl2018dermamnistham10000,codella2019dermamnistskin} for 7-way skin lesion diagnosis in dermoscopy images, 
DeepDRID~\cite{liu2022retinamnistdeepdrid} for retinal disease grading in fundus photography, 
NCT-CRC~\cite{kather2019pathmnist} for 9-class colorectal cancer histology classification in histopathology images, 
BUSI~\cite{al2020busi} for binary classification of benigh and malignant breast tumors in ultrasound images, 
and Kermany~\cite{kermany2018pneumnist} for pneumonia detection in chest X-ray. 

Results are shown in \cref{fig:reconcls-radar}. \methodname achieves the highest average score across metrics and imaging modalities, with significant improvements of mAP and AUC over PUMIT and MedVAE. These results show that \methodname not merely encodes images in the superficial pixel level, but also retains clinical details necessary for downstream diagnostic tasks.

\begin{figure*}[tb]
    \centering
    \includegraphics[width=\linewidth]{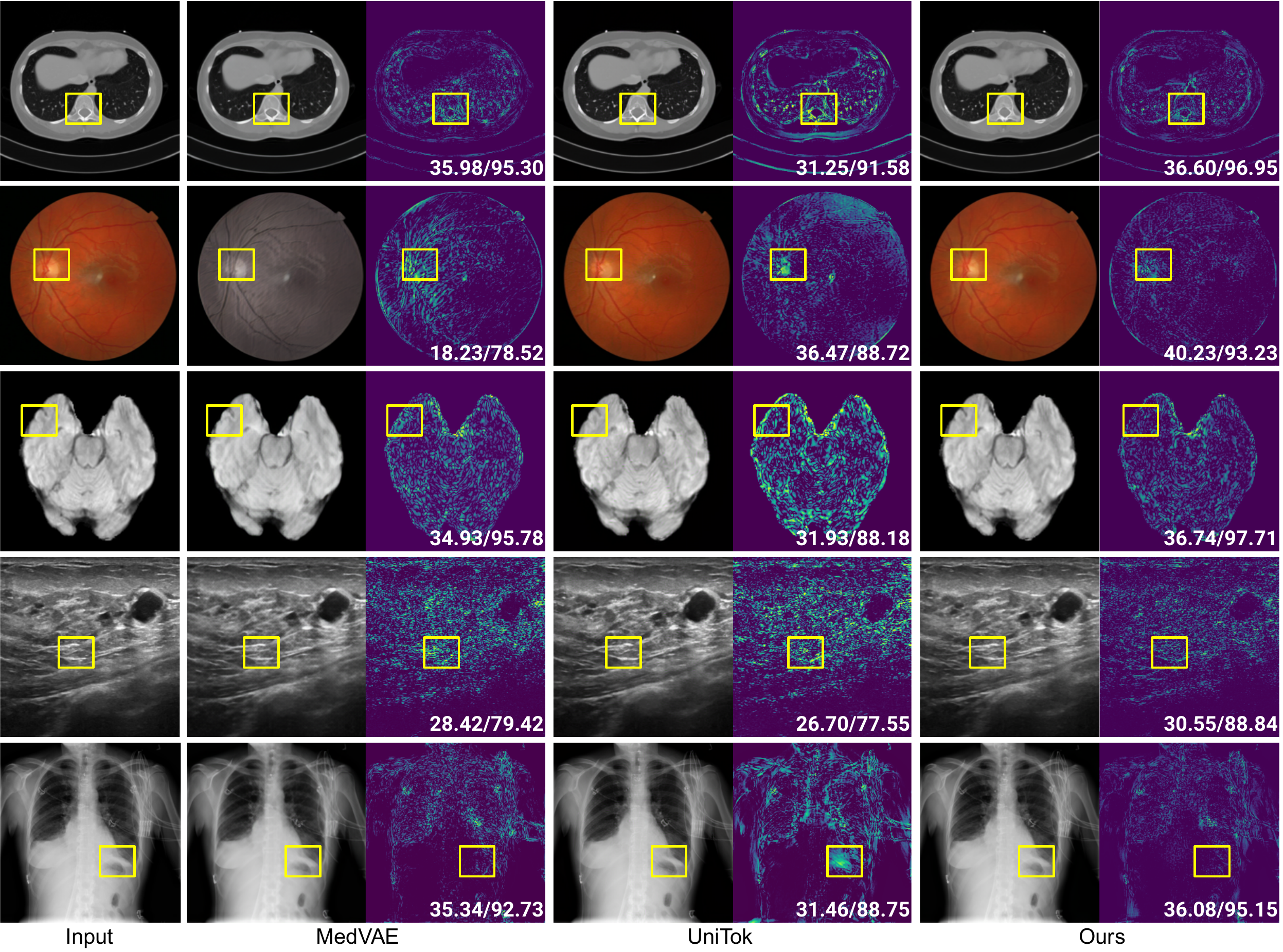}
    \caption{Reconstruction results across multiple imaging modalities. Each reconstructed image is paired with an absolute error map against the input image with PSNR/SSIM. }
    \label{fig:main-recon}
\end{figure*}

\cref{fig:main-recon} visualizes images reconstructed by different tokenizers and corresponding error maps. MedVAE fails to preserve colors due to limited generalizability beyond radiological images, while UniTok discards nuanced details. By contrast, our \methodname consistently preserves fine-grained structures and color fidelity. 
Please refer to 
Appendix~\ref{sec:supp:result} for more results.

\begin{table*}[!t]
\centering
\small
\caption{Downstream image classification performance (mAP / AUC) with linear probing setup. The best results are highlighted in \textbf{bold} and the second best results are \underline{underlined}. Values are presented as percentages. 
}
\resizebox{1.0\textwidth}{!}{%
\begin{tabular}{lccccc|c}
\toprule
Models & Dermoscopy & Fundus & Pathology & Ultrasound & X-ray & Avg. \\
\midrule

VQGAN     &35.71/85.97 &41.59/77.33 &72.69/94.57 &73.29/76.35 &91.34/93.32 &62.92/85.51 \\
Emu3-VQ   &30.79/82.88 &38.90/71.71 &42.57/82.75 &82.65/85.30 &92.75/93.29 &57.53/83.19 \\
VAR-VQ   &58.76/94.02 &51.71/\underline{85.53} &90.80/98.31 &87.31/\underline{89.06} &97.56/97.79 &77.23/\underline{92.94} \\
TokenFlow 
&61.78/93.50 &52.07/83.77 &95.21/99.23 &\textbf{88.19}/88.12 &\underline{97.69}/\underline{98.03} &78.99/92.53 \\
UniTok    
&\underline{66.16}/\underline{94.60} 
&\underline{55.94}/85.05 
&\underline{96.63}/\underline{99.49} 
&87.34/88.60 
&95.98/96.84 &\underline{80.41}/92.92 \\
\cmidrule{1-7}
PUMIT
&23.64/71.92 &36.60/72.87 &81.52/96.50 &68.81/73.67 &88.80/91.64 &59.87/81.31 \\
MedVAE    
&37.66/85.26 &39.31/75.29 &48.02/84.85 &77.74/82.36 &95.41/95.47 &59.54/84.64 \\
\rowcolor{gray!20} \methodname (ours)      &\textbf{71.52}/\textbf{95.60} &\textbf{56.41}/\textbf{86.88} &\textbf{96.88}/\textbf{99.60} &\underline{87.45}/\textbf{89.07} &\textbf{99.08}/\textbf{99.19} &\textbf{82.27}/\textbf{94.07} \\
\bottomrule
\label{tab:main-cls}
\end{tabular}
}
\vspace{-10pt}
\end{table*}

\subsection{High-Level Encoding Evaluation}\label{sec:exp-cls}

To assess whether an image tokenizer encodes high-level clinical semantics in the latent space, we adopt a linear-probing~\cite{alain2016linearprobe} protocol on a suite of medical image classification datasets used in Sec.~\ref{sec:exp-recon}, including ISIC2018 for dermoscopy, DeepDRID for fundus photography, NCT-CRC for pathology images, BUSI for ultrasound, and Kermany for X-ray. Specifically, given a dataset, we freeze each tokenizer and append a linear layer to its encoder, training the linear layer to convergence on the image classification task and report the performance in terms of mAP and AUC on the corresponding test set. 

Results are presented in \cref{tab:main-cls}. Models optimized purely for image reconstruction (\eg, Emu3-VQ, PUMIT) achieve reasonable performance on relatively easier tasks like detection of pneumonia in X-ray images. However, they 
degrade on tasks where fine-grained clinical semantics are required for nuanced classification, \eg, retinal disease grading in fundus photographs. General-domain unified tokenizers like TokenFlow and UniTok show improved but limited performance. By contrast, our \methodname encodes rich clinical-relevant semantics and delivers the best overall classification performance, showing that rich semantic information is embedded in \methodname's unified latent space.

\begin{table}[t]
\centering
\caption{Medical image synthesis performance comparison (gFID; lower is better) across three different synthesis tasks.}
\small
\begin{tabular}{lcccr}
\toprule
Models & Imaging modality & Skin disease & Avg. \\
\midrule
LlamaGen\textsubscript{VQGAN}         & 130.93\footnotesize{$\pm$3.58} & 129.02\footnotesize{$\pm$2.54} & 129.97 \\
LlamaGen\textsubscript{UniTok}           &  80.71\footnotesize{$\pm$3.18} &  97.82\footnotesize{$\pm$1.09} &  89.27 \\
LlamaGen\textsubscript{\methodname w/o S1}   &  \underline{80.58}\footnotesize{$\pm$1.45} &  \underline{93.70}\footnotesize{$\pm$1.00} &  \underline{87.14} \\
\rowcolor{gray!20} LlamaGen\textsubscript{\methodname}          &  \textbf{76.78}\footnotesize{$\pm$1.91} &  \textbf{90.95}\footnotesize{$\pm$1.02} &  \textbf{83.87} \\
\bottomrule
\end{tabular}
\label{tab:main-gen}
\end{table}

\begin{figure*}[t]
    \centering
    \includegraphics[width=\linewidth]{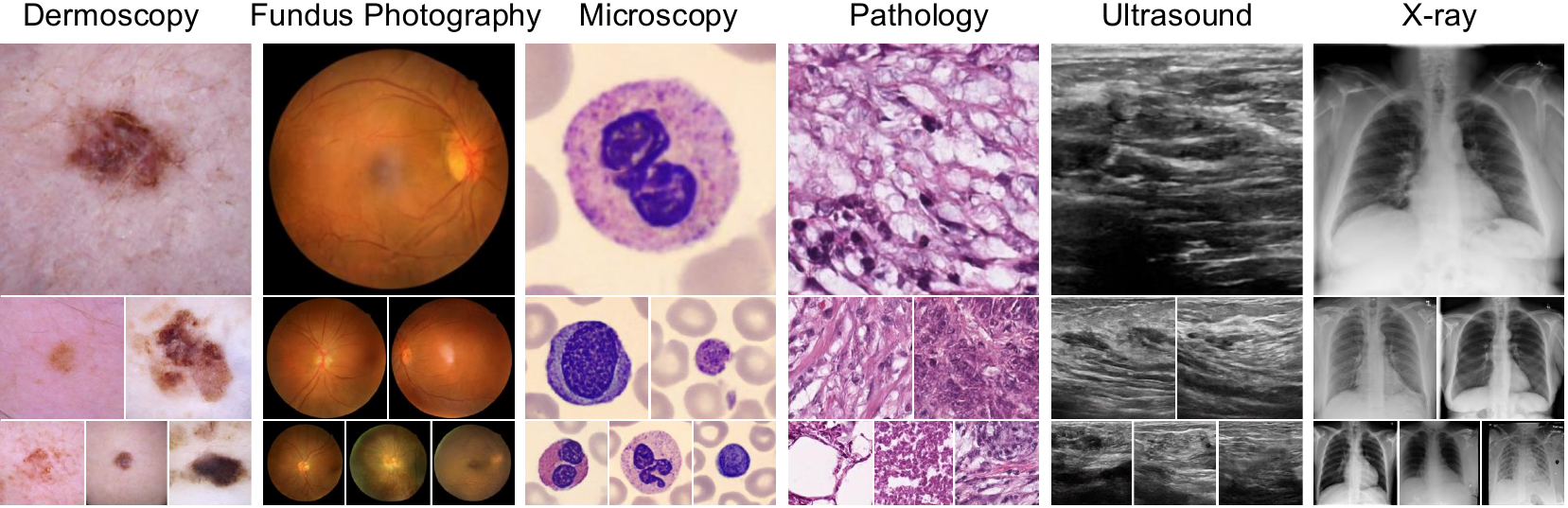}
    \caption{Modality-conditioned synthesized image examples produced by our LlamaGen\textsubscript{\methodname}.}
    \label{fig:main-gen}
\end{figure*}

\subsection{Medical Image Synthesis}\label{sec:exp-syn}
In this part, we evaluate two sub-tasks of conditioned medical image synthesis, \ie, conditioning the synthesis with imaging modalities and skin diseases. For the modality-conditioned synthesis, we construct the dataset using PBC~\cite{acevedo2020bloodmnist}, BUSI, NIH-ChestXray14~\cite{wang2017chestmnist}, ISIC2018, NCT-CRC, and DeepDRID for trainining and testing downstream AR image synthesis models in generating images given a modality from dermoscopy, fundus photography, microscopy, pathology images, ultrasound, and X-ray. 
For the skin disease-conditioned synthesis, we evaluate models using Derm12345~\cite{yilmaz2024derm12345} dataset in generating dermoscopy images of 5 skin diseases, including melanocytic-benign, melanocytic-malignant, non-melanocytic-benign, non-melanocytic-malignant, and non-melanocytic-indeterminate diseases.

We explore applying unified image tokenizers to image synthesis tasks by incorporating each tokenizer in the LlamaGen~\cite{sun2024llamagen} framework. 
Specifically, we build two LlamaGen models using ``\methodname w/o S1'', a variant of \methodname that is only trained with the textual semantic alignment stage without the visual representation alignment, and \methodname. These two models, denoted by ``LlamaGen\textsubscript{\methodname w/o S1}'' and ``LlamaGen\textsubscript{\methodname}'', respectively, are compared with other LlamaGen variants with different image tokenizers, \ie, ``LlamaGen\textsubscript{VQGAN}'' and ``LlamaGen\textsubscript{UniTok}''. We report the generation Fr\'echet inception distance (gFID) for these models; lower gFID indicates higher fidelity of the synthesized images.

\cref{tab:main-gen} shows that LlamaGen using general-domain tokenizer like VQ-GAN or UniTok struggles with high-quality medical image generation. Notably, LlamaGen \textsubscript{\methodname} achieves the best visual fidelity and diversity. We also note that LlamaGen\textsubscript{\methodname} surpasses LlamaGen\textsubscript{\methodname w/o S1} by a non-trivial margin, indicating that the proposed visual representation alignment provides a more regularized token space that boosts the image synthesis task. 
\cref{fig:main-gen} shows images synthesized by LlamaGen\textsubscript{\methodname} across various modalities, presenting realistic structures and textures of biological tissues and organs. Although \methodname is not trained on microscopy modalities, it still supports realistic synthesis of microscopy images. Please refer to Appendix~\ref{sec:supp:result} for more examples.

\begin{figure}[ht]
    \centering
    \includegraphics[width=\linewidth]{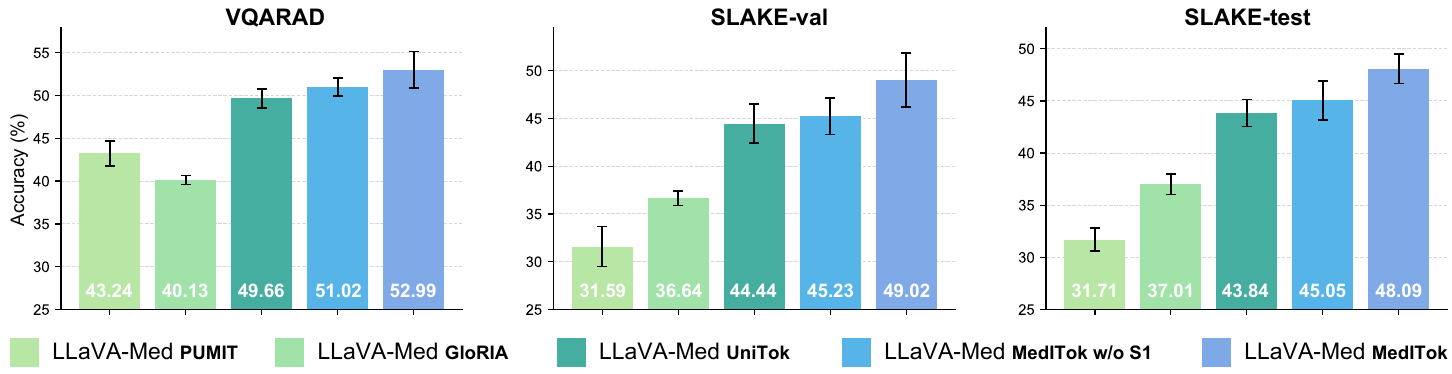}
    \caption{Visual question answering accuracy comparison.}
    \label{fig:main-und}
\end{figure}

\subsection{Medical Image Understanding}\label{sec:exp-vqa}
In this part, we evaluate the medical image understanding capabilities of different image tokenizers via visual question answering (VQA). 
Specifically, we integrate each tokenizer as the image encoder in the LLaVA-Med~\cite{li2023llavamed} framework, yielding the following models: 
LLaVA-Med\textsubscript{GloRIA},
LLaVA-Med\textsubscript{PUMIT}, 
LLaVA-Med\textsubscript{UniTok}, LLaVA-Med\textsubscript{\methodname w/o S1}, 
and LLaVA-Med\textsubscript{\methodname}. For each model, we initialize the language backbone using the released weights of LLaVA-Med, train these models using the image-caption and VQA pairs from a large-scale multimodal medical dataset: PubMedVision~\cite{chen2024huatuogpt}, and evaluate their accuracy on two multimodal medical VQA benchmarks: VQA-RAD~\cite{lau2018vqarad} and SLAKE~\cite{liu2021slake}, including both open-ended and closed-ended questions.

As shown in \cref{fig:main-und}, LLaVA-Med equipped with our final \methodname consistently outperforms other models across all benchmarks. \methodname without the visual representation alignment (``w/o S1'') lags behind the default version, despite being directly trained to align with fine-grained textual semantics. This confirms that the proposed visual representation alignment alleviates the reconstruction-semantic conflict in the unified tokenizer training and improves not only medical image synthesis but also understanding capabilities.  
The underperformance of general-domain tokenizer, UniTok, highlights the importance of domain-specific semantic encoding. A medical semantic encoder, GloRIA~\cite{huang2021gloria}, achieves inferior performance although being trained with pure semantic learning objective, probably due to limited generalization beyond X-ray imaging. PUMIT is trained with multiple medical imaging modalities, yet still performs poorly on medical VQA task, because it mainly encodes low-level spatial details without explicitly learning the high-level semantics. 
These results confirm that \methodname provides effective representations to develop powerful AR models for downstream multimodal medical image understanding tasks.

\begin{table*}[t]
\centering
\caption{Ablation studies of \methodname. ``\#Img'': number of images used in the visual representation alignment stage, ``\#Img-txt'': number of image-text pairs used in the textual semantic alignment stage.}
\small
\resizebox{\textwidth}{!}{
\begin{tabular}{lcccc|ccccc}
\toprule
Idx. & Target Visual Repr. & Target Textual Repr. &\#Img (Stage 1) &\#Img-txt (Stage 2) &PSNR &SSIM &mAP &AUC \\ 
\midrule
\textcolor{orange}{(i)} & -- &BiomedCLIP-T  &-- &1.8M &29.36 &83.67 &77.50 &91.86 \\ 
\textcolor{orange}{(ii)} & BiomedCLIP-V &--  &1.8M &-- &31.38 &84.36 &78.49 &92.25 \\ 
\textcolor{orange}{(iii)} & BiomedCLIP-V &BiomedCLIP-T  &800k &1M &30.03 &84.32 &80.09 &92.64 \\
\midrule
\textcolor{orange}{(iv)} & BiomedCLIP-V &BiomedCLIP-T  &800k &2.4M &29.74 &84.14 &80.28 &92.72 \\
\textcolor{orange}{(v)} & BiomedCLIP-V &BiomedCLIP-T  &2M &2.4M &30.20 &85.50 &82.23 &93.61 \\
\textcolor{orange}{(vi)} & BiomedCLIP-V &BiomedCLIP-T  &33.4M &2.4M &31.74 &88.25 &82.27 &94.07 \\
\midrule
\textcolor{orange}{(vii)} & \mbox{\hspace{10mm}}CLIP-V &--  &800k &-- &30.99 &86.67 &70.80 &89.01 \\
\textcolor{orange}{(viii)} & BiomedCLIP-V &--  &800k &-- &30.00 &83.85 &78.35 &92.23 \\ 
\bottomrule
\end{tabular}
}
\label{tab:abl-basic}
\vspace{-5pt}
\end{table*}

\subsection{Ablation Studies}

\noindent\textbf{Two-Stage Training.}\quad
We first validate the importance of the proposed training framework by ablating the two stages, shown in first three rows of \cref{tab:abl-basic}. The number of images for training are controlled for a fair comparison. 
Row~\textcolor{orange}{(i)} trains the tokenizer without the proposed visual representation alignment stage and directly combines the reconstructive and textual semantic alignment objectives. This version struggles to bridge the gap between two potential conflicting objectives, and shows inferior performance on both low-level encoding (PSNR/SSIM) and high-level encoding (mAP/AUC), as discussed in Secs.~\ref{sec:intro}, \ref{sec:exp-syn}, and~\ref{sec:exp-vqa}. 
Row~\textcolor{orange}{(ii)} only involves visual representation stage and improves the performance, as BiomedCLIP-V encoder provides visual representations that are more compatible for alignment than BiomedCLIP-T. Finally, Row~\textcolor{orange}{(iii)}, which employs our two-stage training, boosts image classification significantly without degrading reconstruction quality, highlighting the effectiveness of visual representation alignment as an intermediate step towards training a unified tokenizer.

\noindent\textbf{Image Data Scaling.}\quad
One notable benefit of our proposed training framework is that it allows effective use of unpaired medical image datasets, which are typically more accessible than image-text data. Rows \textcolor{orange}{(iv)}, \textcolor{orange}{(v)}, and \textcolor{orange}{(vi)} of \cref{tab:abl-basic} illustrate the impact of scaling up the number of unpaired image corpus in the first training stage. Notably, expanding the image data from 800k to 33.4M yields consistent improvements across all metrics, demonstrating the scalability of our proposed approach, allowing it to fully exploit medical image data to enhance both structural fidelity and downstream diagnostic performance. 

\noindent\textbf{Choice of Pre-trained Encoder.}\quad 
Ideally, the choice of pretrained encoders in the proposed framework are flexible, provided that they offer rich semantic representations, \eg, CLIP-family~\cite{radford2021clip,zhang2023biomedclip}. We experiment with two options: the general-domain CLIP~\cite{radford2021clip} and the biomedical-domain BiomedCLIP~\cite{zhang2023biomedclip}. Results in Rows \textcolor{orange}{(vii)} and \textcolor{orange}{(viii)} of \cref{tab:abl-basic} show that, by aligning to BiomedCLIP, \methodname achieves significant improvement in the high-level encoding while maintaining  competitive low-level encoding capabilities in image reconstruction, indicating that domain-specific pre-trained encoders can provide clinically-relevant semantics that benefit downstream tasks. 

Please refer to Appendix~\ref{sec:supp:result} for more experimental results.

\section{Discussion and Conclusion}\label{sec:concl}

In this paper, we propose a novel two-stage training framework that leads to \methodname, a unified medical image tokenizer that encodes both low-level structural details and high-level clinical semantics.
Our framework mitigates the task interference that plagues the joint optimization of reconstructive and semantic objectives by introducing a visual representation alignment stage before the fine-grained textual representation alignment. This initial stage also allows us to effectively leverage large-scale, unpaired medical images, addressing a critical bottleneck in the medical domain, where meticulously paired image-caption data is typically scarce compared to the abundance of unannotated images.

The resulting unified latent space in \methodname shows practical implications for the deployment of advanced downstream medical models. Our extensive evaluations acrosss 4 task families and more than 30 datasets and 9 imaging modalities confirm that \methodname outperforms both general-domain and specialized medical tokenizers. By maintaining high reconstruction fidelity without sacrificing semantic richness, we believe \methodname will serve as a foundational building block for next-generation multimodal AI systems in medical applications.

Despite these advancements, important avenues remain for future exploration. First, \methodname is currently designed for medical image data. While it can be extended to volumetric data (Appendix~\textcolor{eccvblue}{D}), developing a native 3D tokenization architecture to fully capture high-dimensional spatial information is a critical next step. Second, while the training framework uses BiomedCLIP embeddings, one can explore representations that are stronger or more suitable for custom tasks. Finally, as a foundational building block rather than an end-to-end system, executing advanced workflows will require integrating \methodname into broader architectures. 
Please refer to 
Appendix~\textcolor{eccvblue}{E} for more discussion.

\newpage

\setcounter{figure}{0}
\setcounter{table}{0}
\setcounter{equation}{0}
\renewcommand{\thetable}{S\arabic{table}}
\renewcommand{\thefigure}{S\arabic{figure}}
\renewcommand{\theequation}{S\arabic{equation}}

\appendix

\begin{center}
	\bf \LARGE{--- Appendix ---\\}\large{Unified Medical Image Tokenizer for Autoregressive Synthesis and Understanding}
\end{center}

\phantomsection\label{toc:appendix}



\begin{figure*}[thb]
    \centering
    \includegraphics[width=\linewidth]{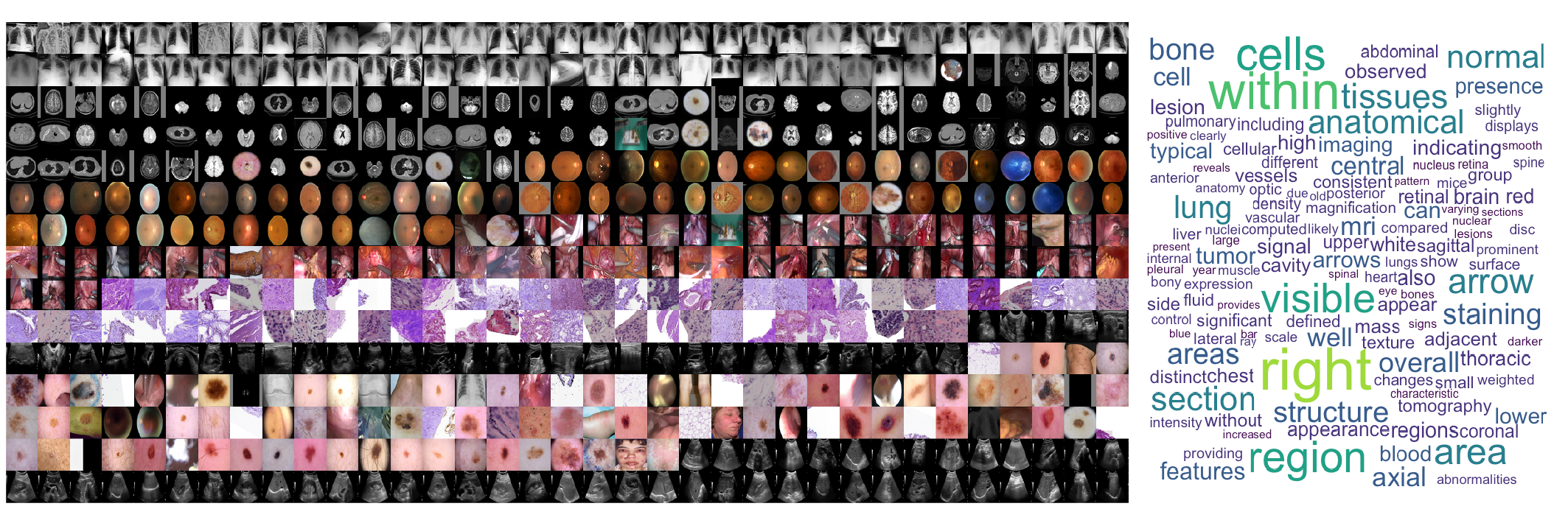}
    \caption{Overview of the training data for \methodname. Left: exemplar images used in the first training stage. Right: word cloud generated from the texts used in the second training stage.}
    \label{fig:supp-teaser}
\end{figure*}

\section*{TL;DR}
This appendix provides a comprehensive detailed account of the data, implementation, and extended analysis for \methodname. We begin by elaborating the construction and quality control of the training dataset (Appendix~\ref{sec:supp:training-dataset}) and the specific protocols for benchmarking datasets (Appendix~\ref{sec:supp:benchmarking-datasets}). Subsequently, we detail the experimental implementation (Appendix~\ref{sec:supp:setup}). We then present extended experimental results and additional qualitative visualizations (Appendix~\ref{sec:supp:result}). Finally, we discuss design choices, comparisons with related work, and societal implications (Appendix~\ref{sec:supp:discussion}).

Note that links provided in the appendix direct exclusively to third-party repositories for reference purposes and strictly adhere to double-blind anonymity.

\section{Training Dataset}
\label{sec:supp:training-dataset}
In this section, we provide a comprehensive overview of the training dataset used in this work, including the collection (Appendix~\ref{sec:supp:training-dataset-data-collection}), preprocessing (Appendix~\ref{sec:supp:training-dataset-data-preprocessing}), and statistics (Appendix~\ref{sec:supp:training-dataset-data-statistics}) of image-only datasets and image-text paired datasets. The construction of this training dataset is pivotal to the success of our proposed \methodname, as it ensures a diverse and high-quality representation of medical images and text descriptions across multiple modalities, anatomical regions, and clinical contexts.

\subsection{Data Collection}
\label{sec:supp:training-dataset-data-collection}
We begin by identifying and collecting medical imaging datasets from over 300 publicly available sources, ensuring broad coverage of imaging modalities and clinical scenarios. Our selection criteria include:
    \textbf{(1)} Appropriate Licensing: We only select datasets with clear licensing, ensuring compliance with data usage policies;
    \textbf{(2)} Clinical Relevance: Only datasets that provide diagnostic-quality images or clinically annotated images were included; 
    and \textbf{(3)} Diversity in Imaging Modalities and Anatomies: We prioritize datasets that collectively cover a wide range of anatomical regions and pathologies.

\subsection{Data Preprocessing}
\label{sec:supp:training-dataset-data-preprocessing}

\subsubsection{Radiology Image Preprocessing.}
\label{sec:supp:training-dataset-data-preprocessing-2d3d}
Radiology images take up a significant portion of our dataset. They are stored in their unique values (\eg, CT images are stored in HU values instead of uint8 pixel intensities) and require modality-specific preprocessing. 

In this part, we describe how radiology images like CT and MRI are preprocessed for the first training stage, \ie, the visual representation alignment stage. We note that this stage is \emph{essentially self-supervised and does not involve textual captions or reports,} and thus can leverage abundant unpaired medical images.


\paragraph{CT image preprocessing.} Each CT image is clipped to the range of $[-1000, 2000]$. To obtain 2D slices from the 3D volume, we extract slices along three orthogonal planes (axial, coronal, and sagittal). 
We then perform an initial quality filtering by retaining CT slices that met all the following criteria: (1) a background ratio (the proportion of pixels with HU values $\leq -1000$) $\leq 0.6$, (2) a valid body ratio (the proportion of pixels with HU values $\geq -300$) $\geq 0.1$, and (3) a pixel intensity standard deviation $< 100$. These criteria ensure the removal of largely empty slices with minimal anatomical content.

Note that, we save the CT images in their original HU values without scaling them to the $[0, 255]$ range. By doing so, we can apply various CT window settings on the CT images during model training as a form of data augmentation, as detailed in Appendix~\ref{sec:supp:setup-training}. 

\paragraph{MRI image preprocessing.} The intensities of MRI images are first clipped to the $[0.5^\text{th}, 99.5^\text{th}]$ percentile range, followed by min-max normalization to $[-1, 1]$. The 2D slices are extracted using the same way as CT preprocessing. The initial quality filtering for MRI excludes those slices with mean pixel values $\leq -0.9$ or standard deviation $\leq 0.2$.

\begin{figure}[htbp]
    \centering
    \includegraphics[width=\linewidth]{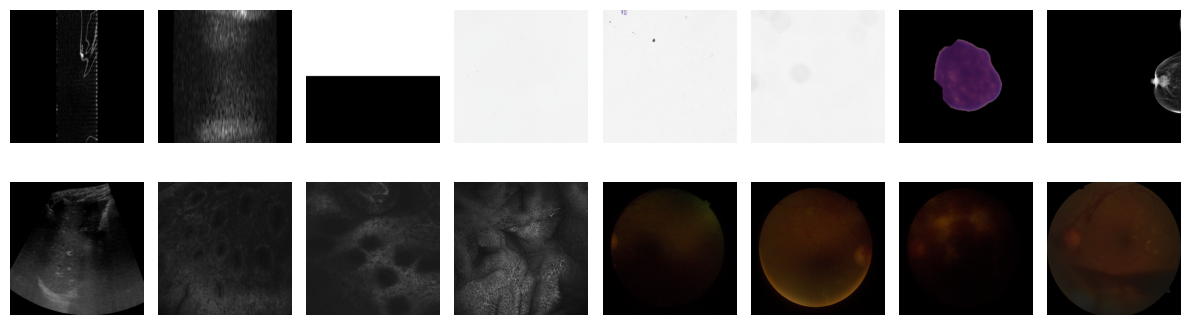}
    \caption{Low-quality images filtered by our quality control pipeline.}
    \label{fig:filtered1}
\end{figure}

\subsubsection{Quality Control.}
\label{sec:supp:training-dataset-data-preprocessing-qc}
Once we obtain all the 2D images, we implement the following process to ensure that only high-quality, clinically relevant images are included in the training dataset:
\begin{itemize}
    \item Dynamic Range Check: Images with pixel intensity ranges below 50 were filtered out to ensure adequate contrast.
    \item Resolution Filtering: Images with a minimum dimension below 128 pixels were excluded to maintain structural integrity.
    \item Information Content Validation: Images with low standard deviation (below~10) in pixel values were discarded.
    \item Palette Limitation Removal: Images with three or fewer unique pixel values were removed.
    \item Relevance Verification: Non-clinical images, such as tables, plots, or irrelevant illustrations, were manually screened and excluded.
\end{itemize}

For instance, the ``Relevance Verification'' is mainly applied on the BIOMEDICA~\cite{lozano2025biomedica} dataset, which originally contains approximately 24,050,423 image-text pairs extracted from biomedical publications. Each image-text pair is tagged with primary and secondary labels. We retain only those pairs with a primary label of ``Clinical Imaging'' and a secondary label matching one of the following: ``computerized tomography,'' ``clinical imaging,'' ``light microscopy,'' ``immunohistochemistry,'' ``endoscopy,'' ``eye,'' ``X-ray radiography,'' ``ultrasound,'' ``magnetic resonance,'' ``brain,'' ``skin lesion,'' and ``mammography''. Image-text pairs tagged with irrelevant secondary labels (\eg, ``scientific illustration,'' ``ambiguous,'' ``plot,'' ``diagram,'' \etc) are all excluded. Such filtering significantly reduces the BIOMEDICA dataset from 24,050,423 to 1,216,529 image-text pairs for use in our experiments.

Following the automated checks described above, we perform a manual quality assessment by randomly sampling 10 images from each dataset for manual visual inspection. If any low-quality outliers are identified, we further examine other images from the corresponding dataset to evaluate overall quality. Finally, we try our best to remove the images that share the same sources with data in our benchmarking datasets in Appendix~\ref{sec:supp:benchmarking-datasets}.

\begin{figure}[htbp]
    \centering
    \includegraphics[width=\linewidth]{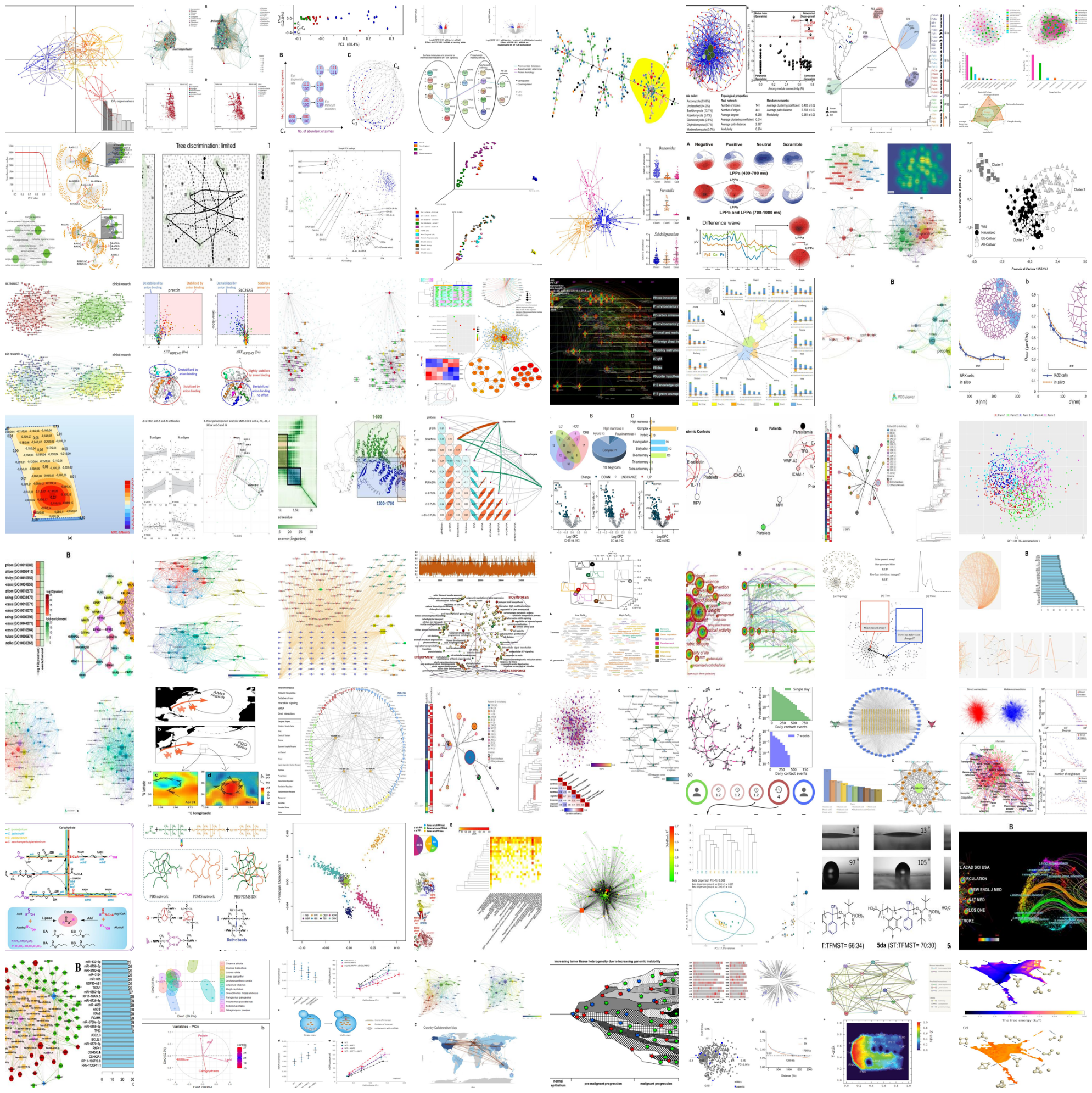}
    \caption{Irrelevant images filtered out by our quality control pipeline.}
    \label{fig:filtered2}
\end{figure}

Fig.~\ref{fig:filtered1} displays some low-quality images detected by the dynamic range check, information content validation, and palette limitation removal. 
For another example, Fig.~\ref{fig:filtered2} shows images that are not tagged as ``clinical imaging'' in the original BIOMEDICA~\cite{lozano2025biomedica} dataset.

\subsection{Data Statistics}\label{sec:supp:training-dataset-data-statistics}

After the collection and the preprocessing, we present detailed sources and image counts of our ``image-only'' dataset, which is used in the first training stage of \methodname, in Tables~\ref{tab:datasets-s1-part1}--\ref{tab:datasets-s1-part4}. The details of the ``image-caption'' dataset, used in the second training stage, are presented in Table~\ref{tab:datasets-s2-part1}.

\begin{blockquote}
\small\emph{Return to: }
\hyperref[toc:appendix]{\textbf{Title Page}}
\end{blockquote}

\section{Benchmarking Datasets}
\label{sec:supp:benchmarking-datasets}

This section outlines the datasets used for evaluating the performance of \methodname across four core task families: low-level encoding (Appendix~\ref{sec:supp:benchmarking-datasets-recon}), high-level encoding (Appendix~\ref{sec:supp:benchmarking-datasets-cls}), medical image synthesis (Appendix~\ref{sec:supp:benchmarking-datasets-syn}), and medical image understanding (Appendix~\ref{sec:supp:benchmarking-datasets-vqa}). We manually validate each dataset to avoid any overlap or data leakage between the training data of \methodname and these benchmark datasets.

\subsection{Low-Level Encoding Evaluation}\label{sec:supp:benchmarking-datasets-recon}
We curated a high-quality evaluation set of 35,736 images spanning 8 imaging modalities and employ image reconstruction task for assessing the low-level encoding capabilities of \methodname. These images are collected from 23 publicly available datasets, as detailed in Table~\ref{tab:datasets-recon-part1}. Importantly, all images used for evaluation are strictly excluded from the training corpus to prevent any overlap. This evaluation set reflects a diverse distribution of anatomical structures, imaging protocols, and clinical contexts, enabling robust testing of image fidelity and structural preservation.

\subsection{High-Level Encoding Evaluation}\label{sec:supp:benchmarking-datasets-cls}
We adopt five subsets to evaluate the semantic encoding quality of the visual tokens produced by different tokenizers. These include: 
\begin{itemize}
\item BUSI~\cite{al2020busi} (ultrasound): binary classification of benign vs. malignant tumors.
    \item ISIC2018~\cite{tschandl2018dermamnistham10000,codella2019dermamnistskin} (dermoscopy): 7-way classification of skin lesions.
    \item NCT-CRC~\cite{kather2019pathmnist} (pathology): 9-class colorectal cancer tissue types.
    \item Kermany~\cite{kermany2018pneumnist} (X-ray): pneumonia detection in chest radiographs.
    \item DeepDRID~\cite{liu2022retinamnistdeepdrid} (fundus): diabetic retinopathy grading.
\end{itemize}
Images in each benchmark are resized to $256\times 256$ resolution for consistency with the different tokenizer input requirements. These tasks collectively test the extent to which the visual tokenizer encodes discriminative, clinically meaningful semantics. Detailed training and test split can be found in Table~\ref{tab:datasets-downstream}.

\subsection{Image Synthesis}\label{sec:supp:benchmarking-datasets-syn}
To evaluate the generative capability of downstream autoregressive models built on top of \methodname, we conduct experiments on modality-conditioned image synthesis and skin disease-conditioned image synthesis. 

For the former, we use six image classification datasets, including PBC~\cite{acevedo2020bloodmnist} for microscopy, BUSI~\cite{al2020busi} for ultrasound, NIH-ChestXray14~\cite{wang2017chestmnist} for chest X-ray, ISIC2018~\cite{tschandl2018dermamnistham10000,codella2019dermamnistskin} for dermoscopy, NCT-CRC~\cite{kather2019pathmnist} for pathology images, and DeepDRID~\cite{liu2022retinamnistdeepdrid} for fundus photography. For the latter, we use Derm12345~\cite{yilmaz2024derm12345}, a multi-source dermatoscopic skin-lesion dataset comprising 12,345 dermoscopy images from 1,627 patients with expert dermatologist consensus annotations, and use the five provided super-classes as the conditions for image synthesis, including melanocytic-benign, melanocytic-malignant, non-melanocytic-benign, non-melanocytic-malignant, and non-melanocytic-indeterminate.

We gather the training partition of these subsets with their imaging modality labels or skin disease labels to construct the training data for the downstream medical image synthesis models, which are trained to generate images conditioned on these labels.

\subsection{Visual Question Answering}\label{sec:supp:benchmarking-datasets-vqa}
To test the utility of different visual tokenizers for medical image interpretation in multimodal settings, we benchmark on two widely adopted datasets for visual question answering (VQA) task: (1) VQA-RAD~\cite{lau2018vqarad}: A radiology-specific VQA dataset with natural language questions and answers grounded in diagnostic images. We use its test set containing 451 question-answer pairs for evaluation. (2) SLAKE~\cite{liu2021slake}: A multi-modal, bilingual medical VQA benchmark with more diverse imaging modalities and question types. The validation set (SLAKE-val) with 2,094 questions and test set (SLAKE-test) with 2,099 questions are adopted in our experiments. 

To train vision-language model for medical image interpretation (\ie, LLaVA-Med~\cite{li2023llavamed} variants with different visual tokenizers as the image encoder), we use the PubMedVision~\cite{chen2024huatuogpt} dataset, which consists of high-quality image-question-answer triplets derived from medical publications. All VQA benchmarks are held out from the training set to ensure fair and unbiased evaluation.

\begin{blockquote}
\small\emph{Return to: }
\hyperref[toc:appendix]{\textbf{Title Page}}
\end{blockquote}

\section{Experimental Setups}\label{sec:supp:setup}
In this section, we first describe the detailed implementation and training setup of \methodname (Appendix~\ref{sec:supp:setup-impl}) and its downstream applications (Appendix~\ref{sec:supp:setup-downstream}) on the aforementioned tasks. 

\subsection{Implementation Details}\label{sec:supp:setup-impl}
\paragraph{Architecture.}
\methodname consists of a ViTamin-based~\cite{chen2024vitamin} image encoder and decoder, with a multi-codebook vector quantizer~\cite{ma2025unitok} in the bottleneck. The encoder produces a 2D grid of latent representations, which are discretized using 8 parallel codebooks, each with 4,096 eight-dimensional vectors, resulting in a total vocabulary size of 32,768. The decoder reconstructs the image from quantized latent vectors. 

\paragraph{Training of \methodname. }\label{sec:supp:setup-training}
Both training stages (\ie, visual representation alignment, and textual semantic alignment) share the same reconstruction loss defined as follows:
\begin{align}
    \mathcal{L}_\mathrm{rec}(\hat{\mat{x}}, \mat{x}, \mat{z}_\mathrm{q}, \mat{z}) &= \mathcal{L}_\mathrm{img}(\hat{\mat{x}},  \mat{x}) 
    + \lambda_\mathrm{vq} \mathcal{L}_\mathrm{vq}(\mat{z}_\mathrm{q}, \mat{z}),\\
    \mathcal{L}_\mathrm{img}(\hat{\mat{x}},  \mat{x}) &= 
    \Vert\hat{\mat{x}}-  \mat{x}\Vert_2^2 
    + \lambda_\mathrm{adv}\mathcal{L}_\mathrm{adv}(\hat{\mat{x}}, \mat{x})  \notag\\
    &+ \lambda_\mathrm{perc}\mathcal{L}_\mathrm{perc}(\hat{\mat{x}}, \mat{x}), \\
    \mathcal{L}_\mathrm{vq}(\mat{z}_\mathrm{q}, \mat{z}) & \!=\! 
    \Vert \mat{z}_\mathrm{q} - \mathrm{sg}[\mat{z}] \Vert_2^2 
    \!+\! \beta\Vert \mathrm{sg}[\mat{z}_\mathrm{q}] - \mat{z} \Vert_2^2,
\end{align}
where $\mathcal{L}_\mathrm{adv}$ denotes the adversarial loss~\cite{esser2021taming}, $\mathcal{L}_\mathrm{perc}$ the perceptual loss~\cite{johnson2016perceptual}, and $\mathcal{L}_\mathrm{vq}$ the vector quantization loss~\cite{van2017vqvae}. ``$\mathrm{sg[\cdot]}$'' denotes the stop-gradient operation. We follow the default setting of VQGAN~\cite{esser2021taming} to set $\lambda_\mathrm{adv}$ as an adaptive weight and fix $\beta=0.25$, $\lambda_\mathrm{perc}=1$, and $\lambda_\mathrm{vq}=1$. The discriminator involved in computing $\mathcal{L}_\mathrm{adv}$ adopts the DINOv2~\cite{oquab2023dinov2} architecture. We use the AdamW~\cite{loshchilov2018adamw} optimizers for both \methodname and the discriminator, with betas of $(0.9, 0.95)$ and a weight decay of $0.02$ for \methodname, and $(0.5, 0.9)$ and $0.2$ for the discriminator. The learning rate is initialized at $5\times10^{-4}$ and decayed to $5\times10^{-5}$ via cosine annealing; for the discriminator, it starts at $2\times10^{-5}$ and decays to $2\times10^{-6}$. The two-stage full-data training took approximately 4 days on 8 NVIDIA H100 GPUs. 

We employ random resized cropping, random image flipping, random image rotation for data augmentation in the first training stage. For CT image input in HU values, we further introduce \textbf{CT windowing augmentation}, which randomly applies the following windows on the HU values: full window ($[-1000, 2000]$ HU, probability $p=0.2$), common window ($[-1000, 1000]$ HU, $p=0.3$), soft tissue window ($[-150, 250]$ HU, $p=0.3$), lung window ($[-1400, 200]$ HU, $p=0.15$), and bone window ($[-500, 1300]$ HU, $p=0.05$).

\subsection{Downstream Tasks}\label{sec:supp:setup-downstream}

\paragraph{Medical image classification. }
For classification tasks, we evaluate the discriminative power of the learned visual tokens through a linear probing protocol~\cite{alain2016linearprobe}. Specifically, for a pretrained visual tokenizer (\eg, \methodname), we only use its image encoder and quantizer, keep them frozen, and append a single linear layer on top of the quantizer. Given an image, the image encoder produces the continuous feature maps, which are then discretized to a grid of visual tokens and are flattened to feed the linear layer for image classification. 
The linear layer is trained using the Adam~\cite{kingma2014adam} optimizer with a learning rate of $10^{-4}$ and a batch size of 128. Since the tokens produced by different tokenizers lead to different convergence speed for the linear layer, we train each linear layer until convergence  and report the peak performance for a fair comparison.

\paragraph{Medical image synthesis. } For image synthesis, we integrate the visual tokenizer with LlamaGen-B~\cite{sun2024llamagen}, an autoregressive model designed for image generation, with 12 transformer layers, 12 heads, and 768 token dimension. 
We first tokenize each training image, producing a discrete token sequence. Then, LlamaGen is trained to autoregressively predict the token sequence conditioned on a modality label token. LlamaGen models are optimized using AdamW~\cite{loshchilov2018adamw} with betas of $(0.9, 0.95)$, a weight decay of $0.05$, and a learning rate of $10^{-4}$. The models are trained for 200 epochs with a batch size of 128. We do not employ advanced strategy for sampling (\eg, classifier-free guidance) and synthesize images with a temperature parameter of 1.

\paragraph{Medical visual question answering.} For VQA, we adapt LLaVA-Med~\cite{li2023llavamed} by replacing its image encoder with different visual tokenizers, followed by a projector to produce visual embeddings compatible with the pretrained language backbone (\texttt{llava-med-v1.5-mistral-7b}). We follow the staged training procedure of original LLaVA-Med, which includes a pretraining stage for the projector (with all other components frozen) and a fine-tuning stage for the language model using LoRA~\cite{hu2022lora}. The pretraining is conducted on 500k image-caption pairs in PubmedVision~\cite{chen2024huatuogpt} dataset for one epoch with batch size 32, while the fine-tuning takes two epochs on the 100k visual question-answer pairs.

\begin{blockquote}
\small\emph{Return to: }
\hyperref[toc:appendix]{\textbf{Title Page}}
\end{blockquote}

\section{Additional Results}\label{sec:supp:result}
This section compiles extended evidence to complement the main results, including 
additional ablation studies (Appendix~\ref{sec:supp:result-abl}), 
adaptation to 3D medical volumes (Appendix~\ref{sec:supp:result-3d}), additional visual Turing test for medical image synthesis (Appendix~\ref{sec:supp:result-vtt}), 
analyses of the differences between codebooks across training stages (Appendix~\ref{sec:supp:result-codebook}),
additional VQA benchmarking (Appendix~\ref{sec:supp:result-supp-vqa}),
representative failure cases (Appendix~\ref{sec:supp:result-failure}), 
comparison of data scale and inference efficiency (Appendix~\ref{sec:supp:result-efficiency}), and additional visualizations for reconstruction, synthesis, and VQA (Appendix~\ref{sec:supp:result-vis}), including qualitative generative and VQA examples that illustrate behavior beyond aggregate metrics. 

\subsection{Additional Ablation Studies}\label{sec:supp:result-abl}
Due to the large scale of our training data comprising over 33 million pure images and 2 million image-text pairs, it is highly inefficient and computationally prohibitive to perform iterative ablations on the full training data. Therefore, we sampled a relatively smaller subset consisting of 800k pure images and 1 million image-text pairs for ablation study, unless otherwise specified. This subset preserves the modality distribution of the full dataset, allowing us to evaluate design choices efficiently.

We present additional ablation studies in Table~\ref{tab:abl-supp} to further investigate the effectiveness of our data quality control and the proposed training framework. 

\begin{table*}[htp]
\centering
\caption{More ablation studies of \methodname. ``\#Img'': number of images used in the first training stage, ``\#Img-txt'': number of image-text pairs used in the second training stage. ``BiomedCLIP-T (combined)'': textual semantic alignment is combined with the visual representation alignment as one single stage. ``BiomedCLIP-T$^\dagger$'': the BiomedCLIP~\cite{zhang2023biomedclip} text encoder is activated during training. 
}
\vspace{2pt}
\small
\resizebox{\textwidth}{!}{
\begin{tabular}{lccccc|cccc}
\toprule
Idx. & Vision Target Repr. &Text Target Repr. & Vision Repr. Loss &\#Img &\#Img-txt &PSNR &SSIM &mAP &AUC \\ 
\midrule
\textcolor{orange}{(i)} & BiomedCLIP-V &BiomedCLIP-T (combined) & $\lambda_\mathrm{vis}=0.1$ &800k &1M &29.20 &83.22 &81.10 &91.97 \\
\textcolor{orange}{(ii)} & BiomedCLIP-V &BiomedCLIP-T & $\lambda_\mathrm{vis}=0.1$ &800k &1M &30.03 &84.32 &80.09 &92.64 \\
\midrule
\textcolor{orange}{(iii)} & -- &BiomedCLIP-T & $\lambda_\mathrm{vis}=0$ &800k &24M (all~\cite{lozano2025biomedica}) &32.23 &89.36 &57.97 &76.98 \\
\textcolor{orange}{(iv)} & -- &BiomedCLIP-T & $\lambda_\mathrm{vis}=0$ &800k &1M (filtered~\cite{lozano2025biomedica}) &32.55 &89.49 &63.29 &81.68 \\
\midrule
\textcolor{orange}{(v)} &-- &BiomedCLIP-T & $\lambda_\mathrm{vis}=0$ &0 &1.8M &29.36 &83.67 &77.50 &91.86 \\
\textcolor{orange}{(vi)} &BiomedCLIP-V &BiomedCLIP-T & $\lambda_\mathrm{vis}=0.1$ &800k &1M &30.03 &84.32 &80.09 &92.64 \\
\textcolor{orange}{(vii)} &BiomedCLIP-V &BiomedCLIP-T & $\lambda_\mathrm{vis}=1$ &800k &1M &29.99 &83.02 &82.00 &91.81 \\
\midrule
\textcolor{orange}{(viii)} & BiomedCLIP-V &BiomedCLIP-T$^\dagger$ & $\lambda_\mathrm{vis}=0.1$ &33.4M &2.4M &34.03 &91.05 &51.41 &69.84 \\
\textcolor{orange}{(ix)} & BiomedCLIP-V &BiomedCLIP-T & $\lambda_\mathrm{vis}=0.1$ &33.4M &2.4M &31.74 &88.25 &82.27 &94.07 \\
\midrule
\textcolor{orange}{(x)} & BiomedCLIP-V &-- &Cos. sim &800k &-- &30.18 &84.01 &66.19 &85.77 \\
\textcolor{orange}{(xi)} & BiomedCLIP-V &-- &Contrast &800k &-- &30.00 &83.85 &78.35 &92.23 \\ 
\bottomrule
\end{tabular}
}
\label{tab:abl-supp}
\end{table*}

\paragraph{Separating or Combining Two Stages.} In contrast to previous works, we propose incorporating an extra training stage (\eg, visual represenation alignment) in the training of unified visual tokenizer. A natural idea question comes: what about combining this extra stage and the textual semantic alignment stage together into one? That is, in each iteration, we optimize the following loss function:
\begin{equation}
\begin{aligned}
    \mathcal{L} &= \mathcal{L}_\mathrm{rec}(\hat{\mat{x}}, \mat{x}, \mat{z}_\mathrm{q}, \mat{z}) \\
    &+ \lambda_\mathrm{vis} \mathcal{L}_\mathrm{vis}(\mat{z}_\mathrm{q}, f_\mathrm{vis}
    (\mathcal{E}_\mathrm{vis}(\mat{x}))) \\
    &+ \lambda_\mathrm{txt} \mathcal{L}_\mathrm{txt}(\mat{z}_\mathrm{q}, f_\mathrm{txt}
    (\mathcal{E}_\mathrm{txt}(\mat{t}))),
\end{aligned}
\label{eq:combined-loss}
\end{equation}

In Rows \textcolor{orange}{(i)} and \textcolor{orange}{(ii)} of Table~\ref{tab:abl-supp}, we empirically compare combined training and separated training under the same setting. The combined training only slightly improves semantic metrics but significantly degrades reconstruction quality. This may be attributed to the dominance of semantic alignment objectives, which in turn escalates the inherent conflicts between reconstruction (low-level) and semantic (high-level) alignment objectives. In contrast, the separate training transits from a reconstruction-based tokenizer to a unified tokenizer more smoothly, improving joint optimization of reconstruction learning and semantic learning.

We also note that separating two stages provides more flexibility, particularly when training with significantly imbalanced data collections in the medical domain, where unlabeled images are far more abundant than image-text pairs (14x in our final training set). A staged design allows us to exploit such imbalanced data effectively and provides engineering flexibility for making modifications to the pretrained encoders (\eg, adding trainable parameters), while avoiding potential gradient issues caused by heterogeneous batches. 

Therefore, we opt for the separate training over the combined training for implementing our framework.

\paragraph{Data Quality Control. } Rows \textcolor{orange}{(iii)} and \textcolor{orange}{(iv)} of Table~\ref{tab:abl-supp} presents the result from our pilot study to evaluate the effectiveness of our data quality control pipeline. We pretrain \methodname with pure reconstruction objective in the first training stage, and continue the second training stage on the BIOMEDICA~\cite{lozano2025biomedica} dataset. 

Specifically, in Row \textcolor{orange}{(iii)}, we adopt all 24M image-text pairs in this dataset, while in Row \textcolor{orange}{(iv)}, we use a much smaller subset with approximately 1M pairs, as described in Appendix~\ref{sec:supp:training-dataset-data-preprocessing-qc}. 
Surprisingly, despite the significant reduction in the training dataset size, the tokenizer in Row~\textcolor{orange}{(iv)} exhibits much stronger low-level and high-level encoding capabilities, compared to the one in Row~\textcolor{orange}{(iii)}. This highlights the importance of data quality control in training a powerful visual tokenizer\footnote{We note that this filtering was tailored to downstream tasks that mainly involve clinical images, and that other image types (\eg, tables, plots, and non-clinical images) in BIOMEDICA remain highly valuable for applications that require table understanding or scientific figure interpretation.}.

\paragraph{$\lambda_\mathrm{vis}$ Balancing Reconstruction and Contrastive Learning.} In Rows \textcolor{orange}{(v)}, \textcolor{orange}{(vi)}, and \textcolor{orange}{(vii)}, we explores the effect of different magnitude for the visual representation alignment in the first training stage by varying $\lambda_\mathrm{vis}$ in Eq.~\ref{eq:vq-loss}. 
By setting a light semantic constraint ($\lambda_\mathrm{vis}=0.1$), we observe an improvement across three metrics (PSNR, SSIM, and AUC) while maintaining competitive mAP, and we fix this factor in other experiments.

\paragraph{Freezing the Pretrained Text Encoder.} In Row \textcolor{orange}{(viii)}, we investigate the impact of unfreezing the BiomedCLIP text encoder during the second stage. Although this introduces learnable capacity into the text encoder, it disrupts the stability and alignment of the token space, leading to a trade-off: improved reconstruction metrics but severely degraded downstream classification, compared to the results in Row \textcolor{orange}{(ix)}. This suggests that freezing the pretrained textual backbone acts as an anchor, preserving the semantic information necessary for clinical interpretation.

\paragraph{Visual Representation Alignment Objective.}
We explore two alignment objectives for training \methodname: contrastive learning and cosine similarity (inspired by \cite{yao2025reconstruction}). Comparing Rows~\textcolor{orange}{(x)} and~\textcolor{orange}{(xi)}, we observe that using cosine-similarity loss yields only marginal gains in PSNR but substantially degrads downstream classification, whereas the contrastive objective produces a more discriminative token space, improving both fine-grained classification and maintaining reconstruction quality. 

\begin{table}[htp]
\centering
\caption{Additional evaluation on 3D datasets.}
\small
\begin{tabular}{lccccc}
\toprule
Models &rFID &PSNR &SSIM &mAP &AUC \\ 
\midrule
MedVAE & 20.38 & \textbf{34.21} & \textbf{89.98} & 76.04 & 94.77 \\
UniTok & 6.89 & 31.08 & 86.16 & \underline{83.25} & \underline{96.15} \\
\methodname & \textbf{4.94} & \underline{33.56} & \underline{89.54} & \textbf{84.00} & \textbf{97.71} \\
\bottomrule
\end{tabular}
\label{tab:ext-3d}
\end{table}

\subsection{Adaptation to 3D Medical Volumes}\label{sec:supp:result-3d}
Three-dimensional data are vital in the medical domain. Our initial milestone targeted a 2D image tokenizer, considering that (1) 2D images cover more medical imaging domains (\eg fundus photography, histopathology, \etc), (2) 2D models provide more flexibility, and (3) computational costs.

However, we note that \methodname can also be applied in 3D medical data. We compare \methodname, UniTok, and MedVAE on two 3D datasets: SLIVER07~\cite{heimann2009silver07} for volume reconstruction and LiTS~\cite{s1-lits,xu2019efficient} for multi-class volume classification of 11 body organs. To adapt these 2D tokenizers to 3D volumes, we employed a slice-based strategy: processing individual slices independently and then aggregating either reconstructed slices (for low-level encoding evaluation) or per-slice features (for high-level encoding evaluation). The results are summarized in Table~\ref{tab:ext-3d}.

\begin{figure*}[tbp]
    \centering
    \includegraphics[width=\linewidth]{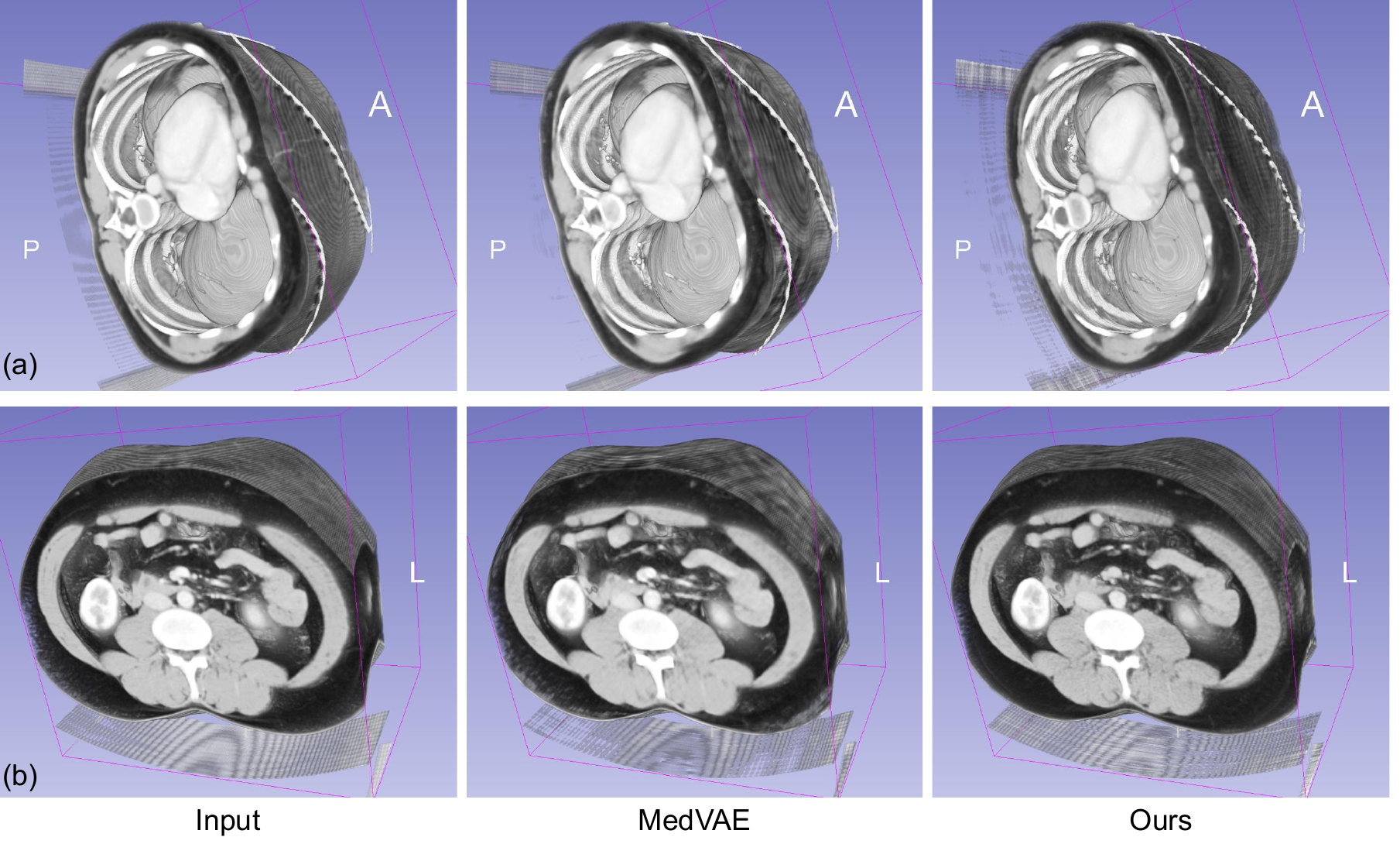}
    \caption{Visualization of 3D reconstruction results.}
    \label{fig:vis-3d}
\end{figure*}

Despite not being trained explicitly on 3D radiology data, \methodname still achieves reconstruction quality comparable to MedVAE which is a \emph{radiology-specialized} visual tokenizer, with notably lower rFID for better visual fidelity and competitive PSNR/SSIM indicating reconstruction accuracy. UniTok encodes visual semantics, yet failing to preserve critical structural details with a significant drop in PSNR and SSIM. More importantly, \methodname significantly outperforms MedVAE on 3D volume classification tasks, proving superior transferable representations in 3D settings. Visualization of 3D reconstruction results are shown in Fig.~\ref{fig:vis-3d}.

\begin{table*}[ht]
\centering
\small
\caption{Visual Turing test on downstream medical image synthesis. We report area under the ROC curve (AUC) for real vs. synthetic discrimination and the fooling rate, with 95\% confidence interval.}
\resizebox{1.0\textwidth}{!}{%
\begin{tabular}{lcccc}
\toprule
\textbf{Model} & \textbf{AUC}\textsubscript{senior} ($\downarrow$)  & \textbf{AUC}\textsubscript{junior} ($\downarrow$) & \textbf{Fooling rate}\textsubscript{senior} ($\uparrow$) & \textbf{Fooling rate}\textsubscript{junior} ($\uparrow$) \\
\midrule
LlamaGen\textsubscript{UniTok}  
& 0.602 (0.430--0.772)  & 0.553 (0.367--0.720) 
& 56.0\% (37.1--73.3\%) & 62.0\% (44.1--80.2\%)  \\
LlamaGen\textsubscript{\methodname} 
& \textbf{0.462} (0.307--0.622) & \textbf{0.403} (0.276--0.597) 
& \textbf{72.0\%} (52.4--85.7\%)  & \textbf{78.0\%} (63.3--90.6\%) \\
\bottomrule
\end{tabular}
\label{tab:vtt}
}
\end{table*}

\subsection{Visual Turing Test}\label{sec:supp:result-vtt}
We conducted a Visual Turing Test on the downstream medical image synthesis task, as a proxy evaluation of the quality of latent space encoded by different tokenizers. Specifically, we compare two autoregressive medical image synthesis models as in 
Sec.~\textcolor{eccvblue}{4.4}: 
(1) LlamaGen\textsubscript{\methodname}, using \methodname as its visual tokenizer; and (2) LlamaGen\textsubscript{UniTok}, using UniTok instead, a state-of-the-art unified visual tokenizer.

We randomly mixed 75 chest X-rays: 25 real, 25 synthesized by LlamaGen\textsubscript{\methodname}, and 25 by LlamaGen\textsubscript{UniTok}, and invited a senior radiologist ($>$ 10 years of experience) and a junior radiologist ($<$ 3 years of experience) to evaluate the images blindly. The radiologists are asked to score the ``realness'' of each image on a continuous 0–1 scale. From these scores, we computed (i) AUC for classifying real versus synthetic images and (ii) ``fooling rate,'' the proportion of synthetic images scored higher than 0.5. As shown in Table~\ref{tab:vtt}, both radiologists had more difficulty distinguishing \methodname-synthesized images from real ones, indicating that \methodname enables a more clinically plausible latent space.

\subsection{Difference between Stages}\label{sec:supp:result-codebook}
We compare the two stages through both performance behavior and the geometry of their learned codebooks. 

\begin{figure}[!htbp]
    \centering
    \begin{minipage}[t]{0.65\linewidth}
        \centering
        \includegraphics[width=\linewidth]{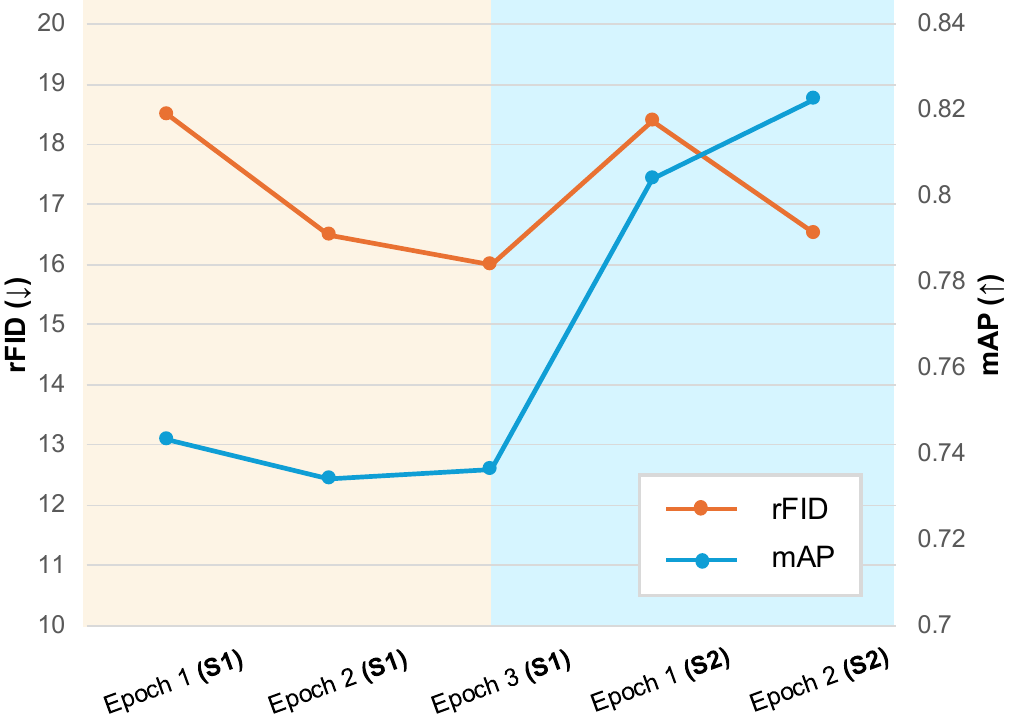}
        \captionof{figure}{Training dynamics of \methodname, where rFID on the reconstruction test set and mAP on the classification test set are reported for checkpoints from three Stage-1 (S1) epochs followed by two Stage-2 (S2) epochs.}
        \label{fig:test-curve}
    \end{minipage}\hfill
    \begin{minipage}[t]{0.3\linewidth}
        \centering
        \includegraphics[width=\linewidth]{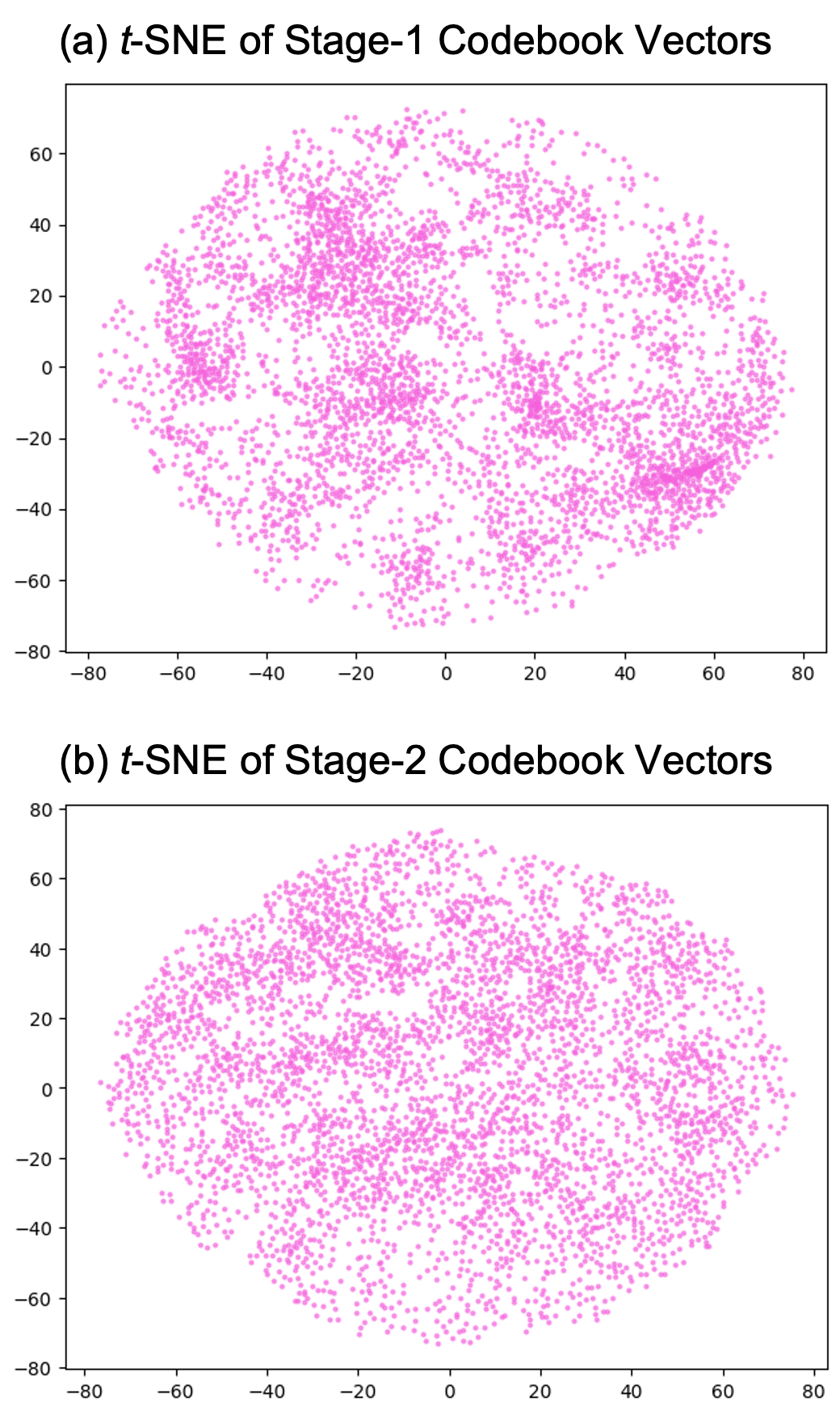}
        \captionof{figure}{$t$-SNE visualization of codebook vectors in two training stages.}
        \label{fig:tsne7}
    \end{minipage}
\end{figure}

Empirically, Fig.~\ref{fig:test-curve} shows the test performance curve. In Stage 1, rFID steadily decreases while mAP remains flat or drifts slightly downward, consistent with a phase that emphasizes reconstructive accuracy over discriminative semantics. When training continues into Stage 2, mAP rises sharply, showing a strong boost in classification performance as semantic constraints are reinforced. rFID exhibits a transient increase at the first epoch in Stage 2 but then returns to a level close to the endpoint of Stage 1, indicating that reconstruction quality is largely preserved. Overall, these dynamics support the design of the two-stage schedule: Stage 1 secures a high-fidelity codebook with light semantic constraint, and Stage 2 further enhances clinical semantics in the latent vectors while retaining structural information encoding.

To understand why, we visualize the codebook vectors with $t$-SNE. As shown in Fig.~\ref{fig:tsne7}, after Stage 2 (strong semantic alignment), the vectors spread more uniformly, pushing features to be well-distributed on the hypersphere, whereas Stage 1 (light semantic constraint) produces visibly clustered pockets.

This is also consistent with known VQ-VAE behavior: without additional pressures, codebooks tend to exhibit codebook collapse~\cite{roy2018codebookcollapse}, yielding concentrated regions in latent space. 
The move toward a more uniform, semantically aligned latent in Stage 2 therefore explains both the stronger interpretive/synthesis performance. This confirms that both stages in training \methodname is important for building the unified token space. Notably, recent work~\cite{yao2025reconstruction} in latent diffusion reaches a congruent conclusion: aligning VAE latents to semantic-rich features promotes generative quality by regularizing the latent geometry, with only limited impact on reconstruction.

\begin{table}[!ht]
\centering
\small
\caption{Visual question answering accuracy comparison.}
\vspace{-5pt}
\begin{tabular}{lrr}
      \toprule
      Models & PMC-VQA test (free-form) \\
      \midrule
      LLaVA-Med 
      & 21.55\scriptsize{$\pm$1.28} \\ 
      LLaVA-Med\textsubscript{UniTok}                       & 22.60\scriptsize{$\pm$1.61} \\ 
      LLaVA-Med\textsubscript{\methodname w/o S1} 
      & 24.74\scriptsize{$\pm$0.95} \\
      \rowcolor{gray!20} LLaVA-Med\textsubscript{\methodname} 
      & \textbf{25.55}\scriptsize{$\pm$1.25} \\
      \bottomrule
\end{tabular}
\label{tab:supp-und}
\end{table}

\subsection{Additional VQA}\label{sec:supp:result-supp-vqa}
We extended our evaluation to the PMC-VQA benchmark~\cite{zhang2024pmcvqa,zhang2023pmcvqa}, which encompasses a diverse range of modalities and pathologies. The original test set consists of multiple-choice questions, and we reformatted the task into free-form question answering to rigorously assess generative capability and prevent models from guessing based on provided options. For instance, a query such as ``What is the color of the articular process in the image? A:Blue B:Red C:Green D:Yellow'' with answer ``C'' was converted into ``What is the color of the articular process in the image?'' with answer ``Green.''

\begin{figure}[!htbp]
    \centering
    \includegraphics[width=\linewidth]{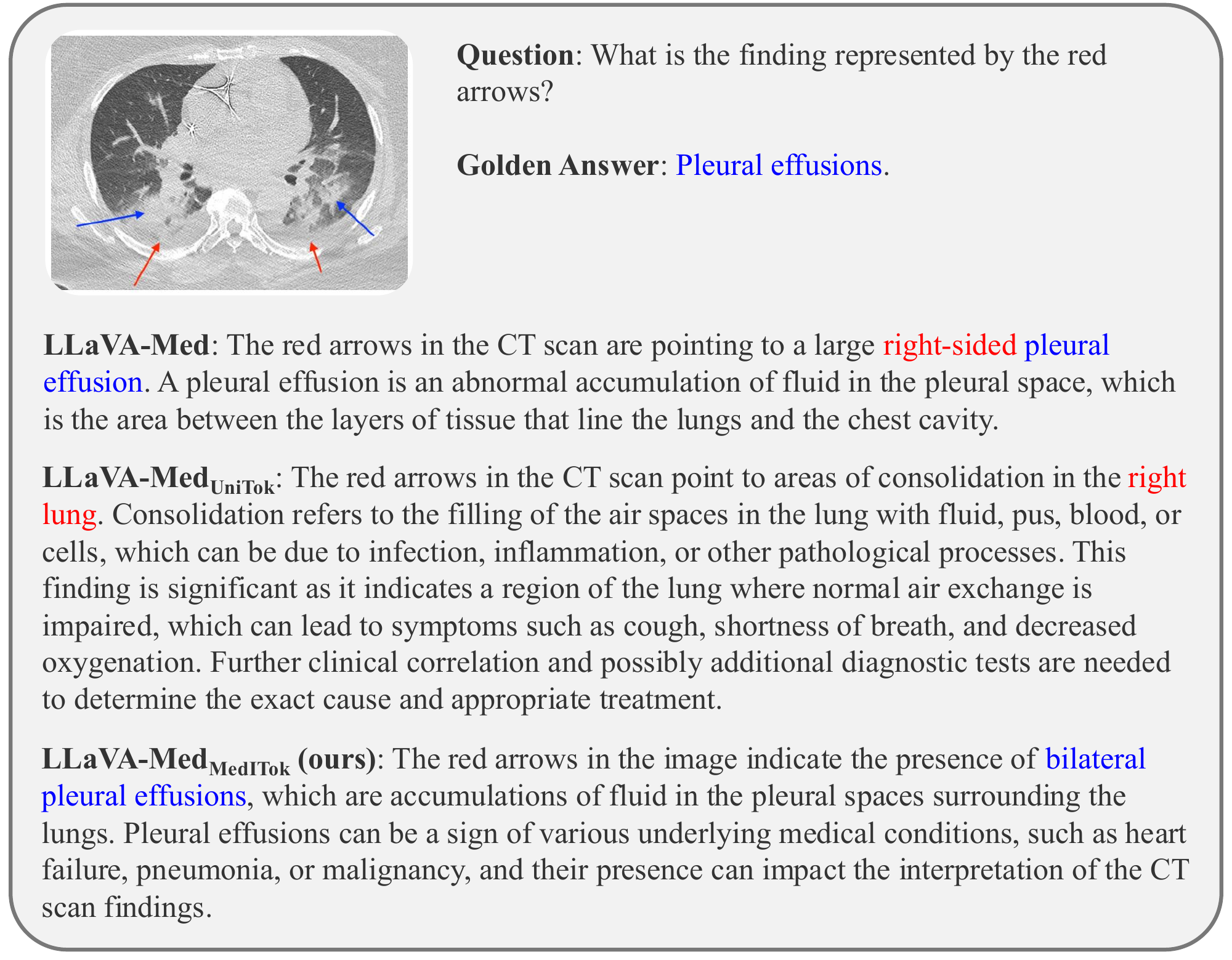}
    \caption{Visual question answering results on a CT scan.}
    \label{fig:vis-vqa4}
\end{figure}

\begin{figure}[!htbp]
    \centering
    \includegraphics[width=\linewidth]{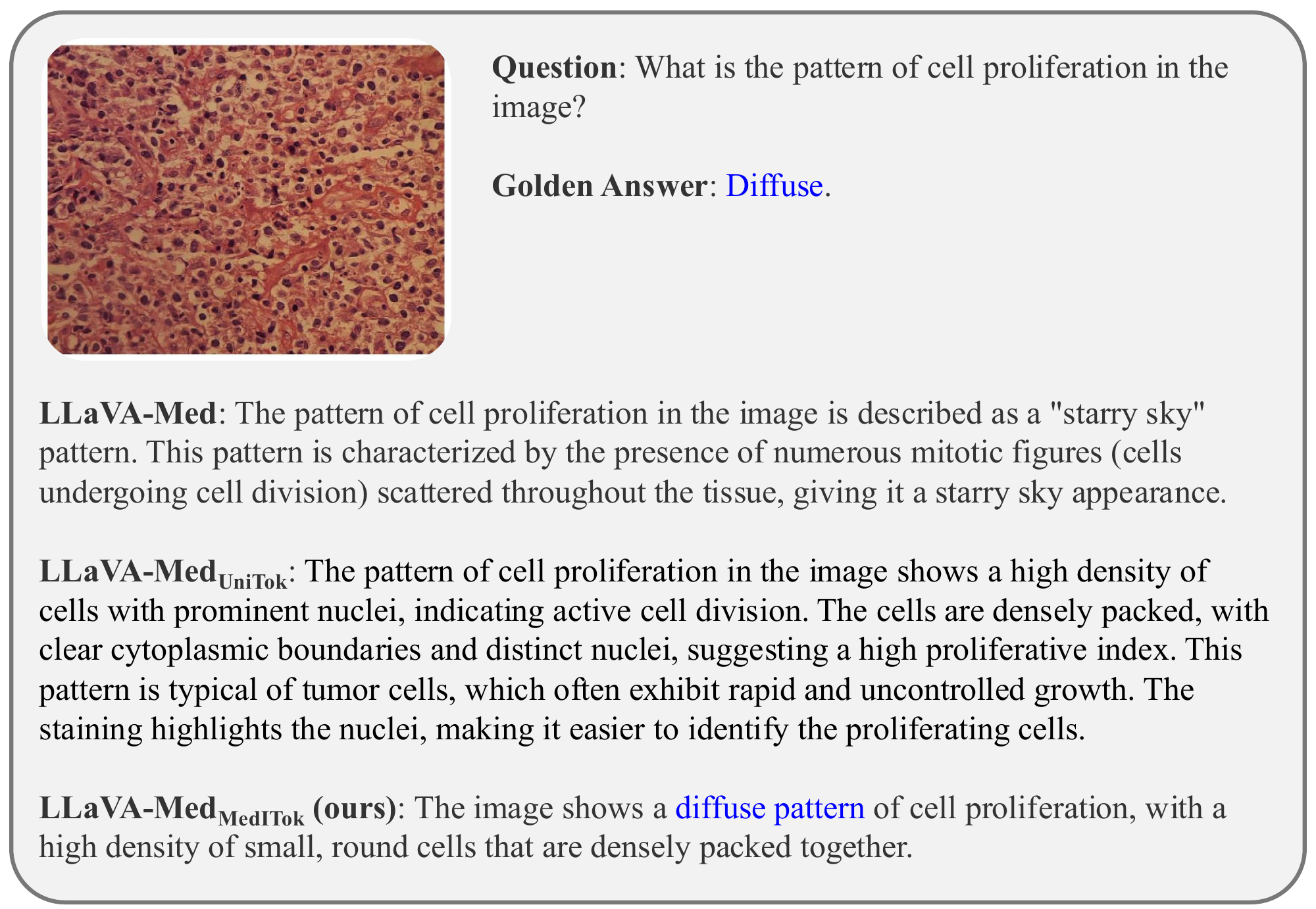}
    \caption{Visual question answering results on a histological image.}
    \label{fig:vis-vqa5}
\end{figure}

\begin{figure*}[!htbp]
    \centering
    \includegraphics[width=\linewidth]{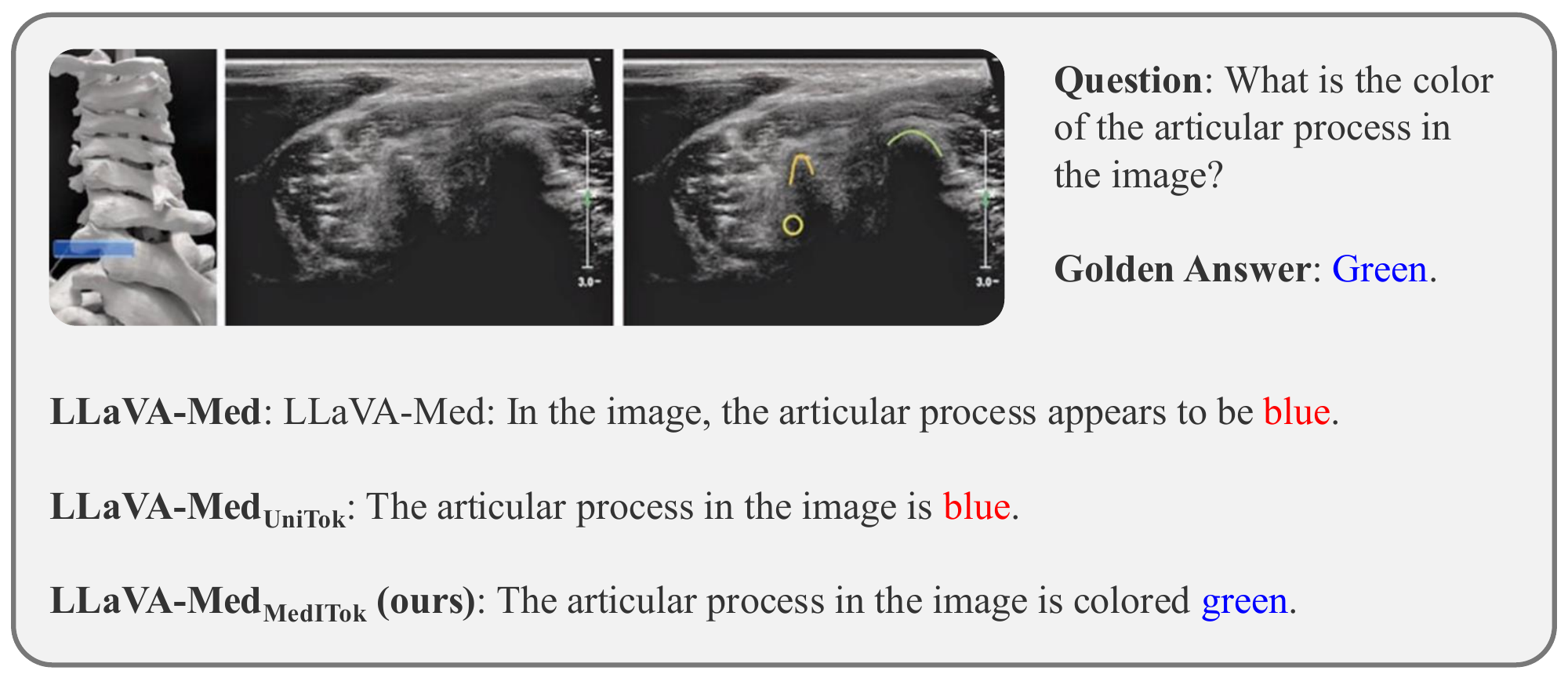}
    \caption{Visual question answering results on ultrasound scans.}
    \label{fig:vis-vqa6}
\end{figure*}

\begin{figure*}[!htbp]
    \centering
    \includegraphics[width=\linewidth]{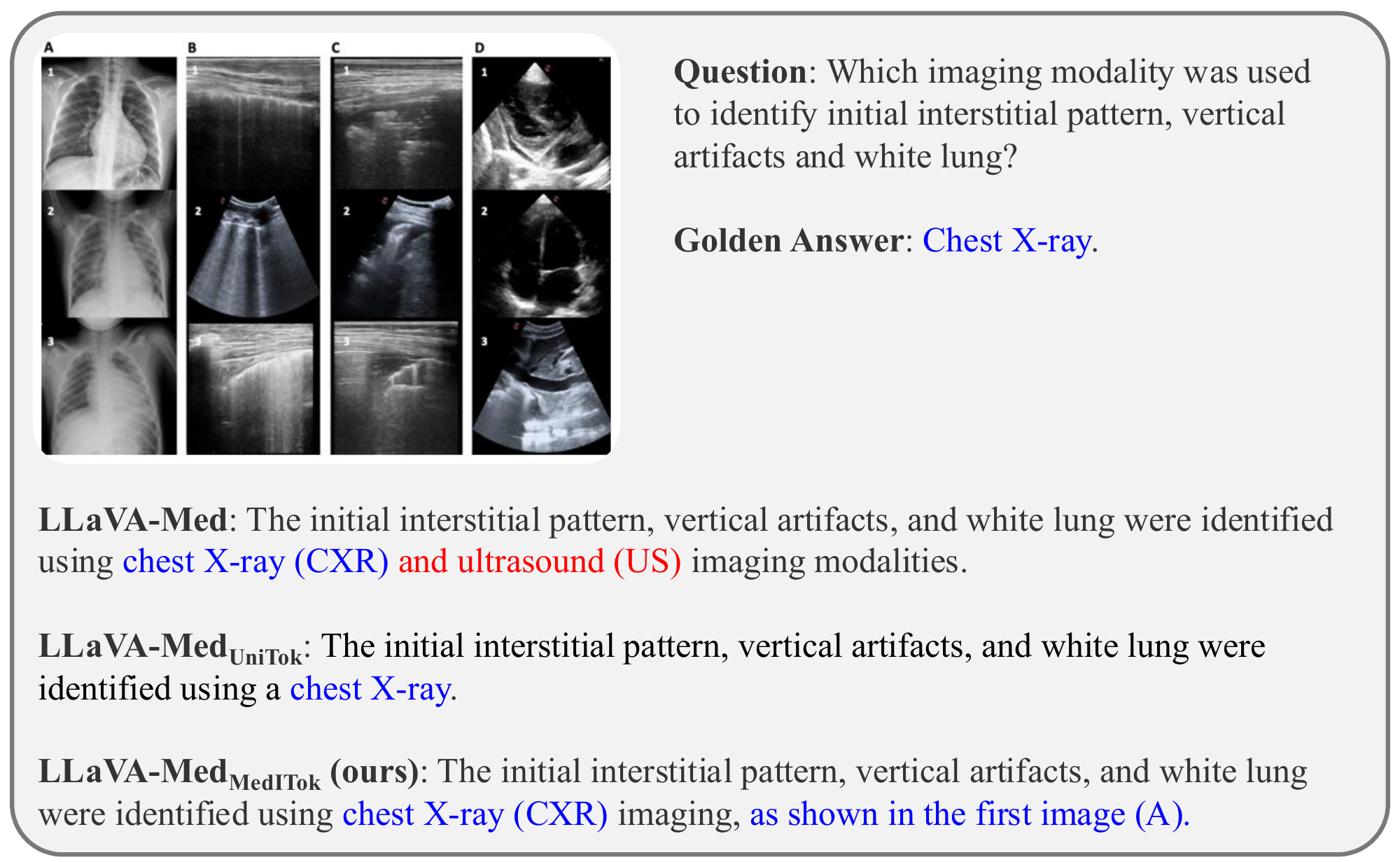}
    \caption{Visual question answering results on multimodal images.}
    \label{fig:vis-vqa7}
\end{figure*}

Quantitative results are presented in Table~\ref{tab:supp-und}, where LLaVA-Med model powered by our proposed tokenizer (LLaVA-Med\textsubscript{\methodname}) achieves the highest accuracy score. Qualitative comparisons, provided in Figs.~\ref{fig:vis-vqa4}, \ref{fig:vis-vqa5}, \ref{fig:vis-vqa6} and \ref{fig:vis-vqa7}, further highlight the model's precision. 
For example, in Fig.~\ref{fig:vis-vqa4} where the golden answer is labeled simply as ``Pleural effusions,'' LLaVA-Med\textsubscript{\methodname} correctly characterizes the finding as ``bilateral pleural effusions,'' whereas other models incorrectly localize the pathology as ``right-sided.'' In Fig.~\ref{fig:vis-vqa6}, the user queries the color of the articular process, which is annotated with a green curve in the right-most ultrasound scan. Baseline models hallucinate the answer ``blue,'' likely  misled by the prominent blue plane in the 3D schematic (left-most figure). In contrast, LLaVA-Med\textsubscript{\methodname} successfully grounds the medical term to the visual features in the ultrasound data, accurately answering ``green.''

\subsection{Failure Cases}\label{sec:supp:result-failure}
\begin{figure*}
    \centering
    \includegraphics[width=0.95\linewidth]{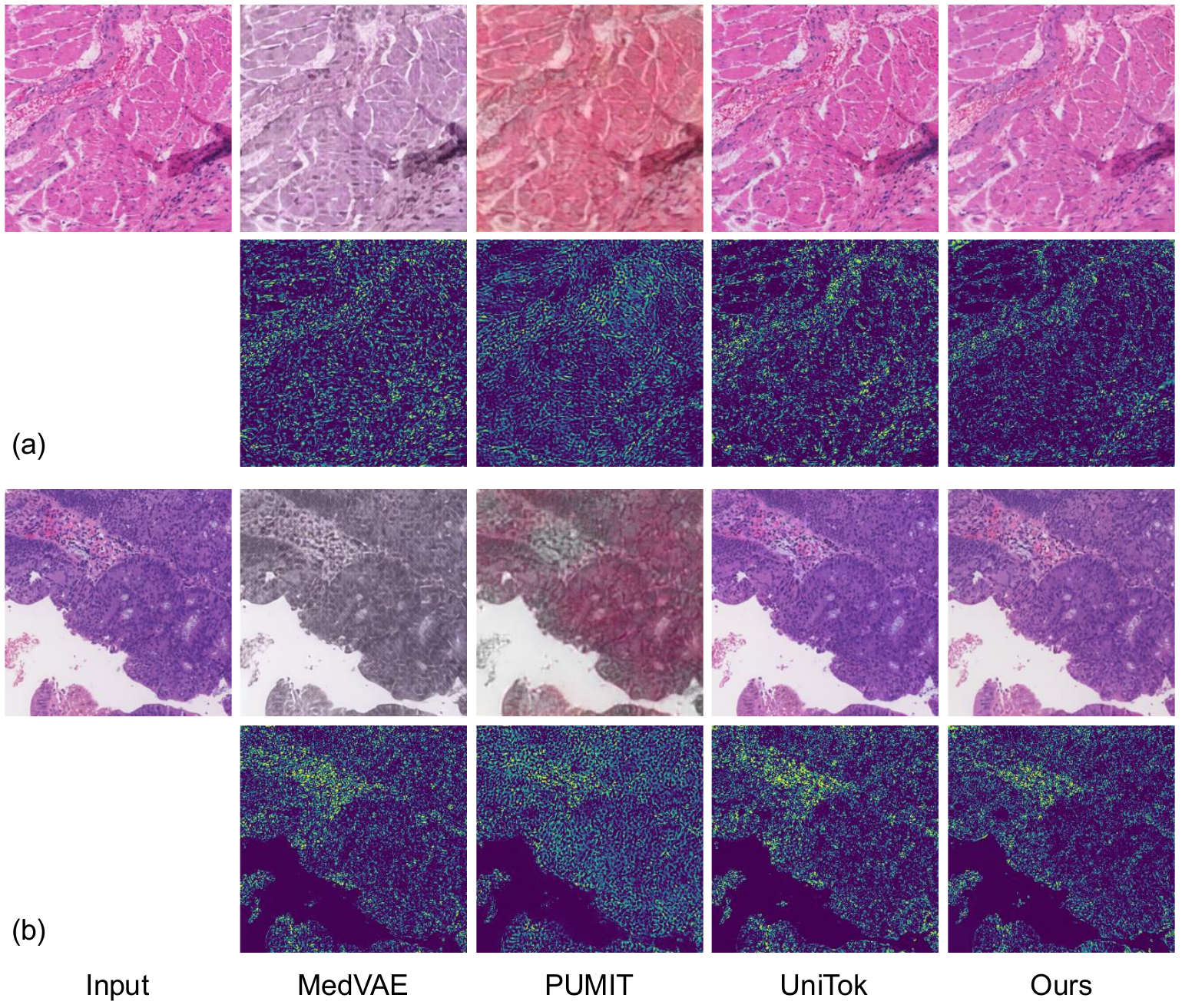}
    \caption{Two failure cases for image reconstruction. Due to the extremely rich details in histopathology images, existing visual tokenizers may still produce lower-fidelity reconstructions.}
    \label{fig:failure}
\end{figure*}

Despite the inspiring performance, \methodname may produce inferior reconstruction for histopathology images due to their rich fine-grained textures and structural complexity. As shown in the ``Patho.'' column of 
Table~\ref{tab:main-recon}, 
all tokenizers struggle with this modality, though \methodname still outperforms existing baselines. This represents a common challenge in histopathology tokenization that warrants future investigation. Qualitative examples for these failure cases are shown in Fig.~\ref{fig:failure}.

\subsection{Efficiency Comparison}\label{sec:supp:result-efficiency}
In Table~\ref{tab:efficiency}, we provide details on the inference GPU memory consumption (GB), and frame-per-second (FPS) throughput across different settings (\eg, B8: batch size~8, R256: resolution~256). \methodname achieves comparable memory consumption and throughput to existing tokenizers while delivering state-of-the-art reconstruction quality and latent representation (Tables~\ref{tab:main-recon} and~\ref{tab:main-cls}), 
showing both efficiency and effectiveness. 

\begin{table*}[ht]
\centering
\small
\caption{Comparison of different models in terms of inference memory usage, and frames per second (FPS).}
\resizebox{1.0\textwidth}{!}{%
\begin{tabular}{l c c c c}
\toprule
\textbf{Model} &  
\textbf{Memory (B16, R256)} & \textbf{Memory (B8, R512)} & 
\textbf{FPS (B16, R256)} & \textbf{FPS (B8, R512)} \\ 
\midrule
VQGAN     & 3.29  & 6.31  & 136.24 & 17.76  \\
PUMIT     & 0.36  & 0.56  & 4440.09 & 1691.37 \\
VAR-VQ    & 4.21  & 7.97  & 171.26 & 40.95  \\
Emu3-VQ   & 41.12 & OOM   & 12.68  & OOM    \\
VAR-VQ    & 4.21  & 7.98  & 171.26 & 40.95  \\
TokenFlow & 7.91  & Not Supported   & 44.15  & Not Supported    \\
MedVAE    & 4.61  & 8.89  & 101.56 & 24.34  \\
\methodname   & 4.69  & 6.75  & 92.81  & 20.63  \\
\bottomrule
\end{tabular}
}
\label{tab:efficiency}
\end{table*}

\subsection{Additional Visualization}\label{sec:supp:result-vis}

\begin{figure*}[!htbp]
    \centering
    \includegraphics[width=\linewidth]{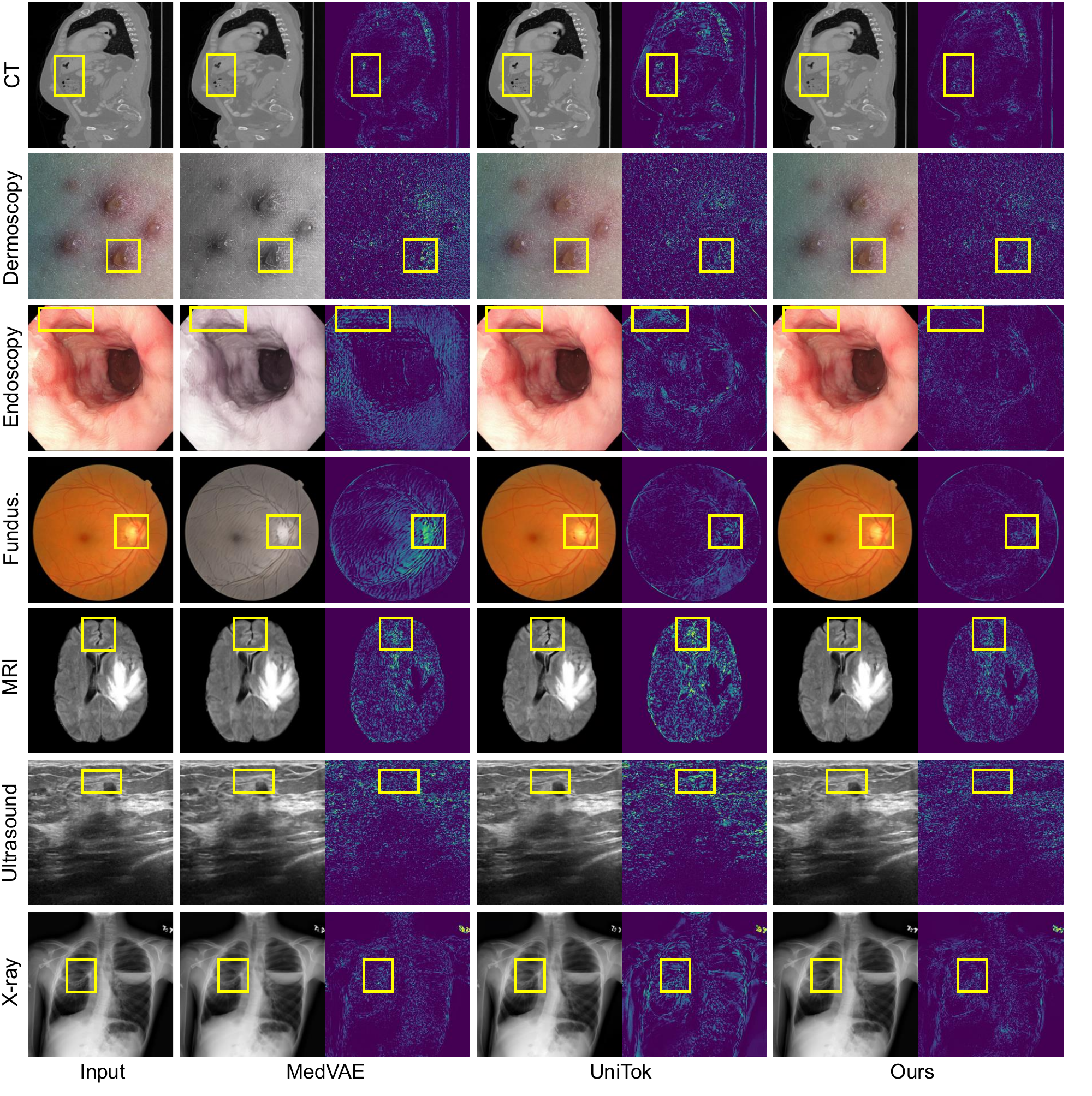}
    \caption{More reconstruction results across multiple imaging modalities. Each reconstructed image is paired with an absolute error map against the input image. Regions of interest are highlighted via yellow bounding boxes.}
    \label{fig:vis-rec-supp}
\end{figure*}

\begin{figure*}[!htbp]
    \centering
    \includegraphics[width=\linewidth]{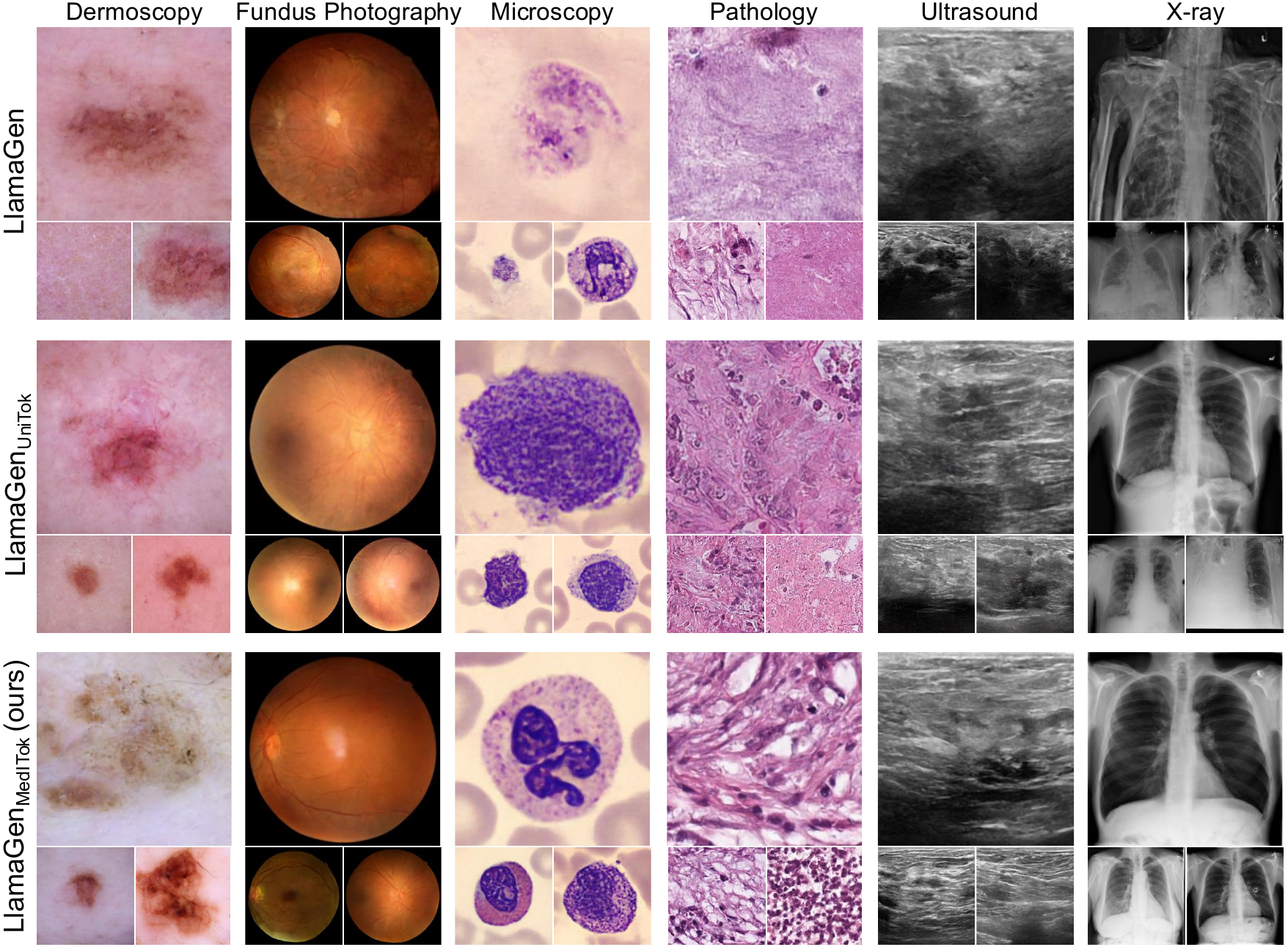}
    \caption{Qualitative comparison of medical images generated by LlamaGen models based on different visual tokenizers.}
    \label{fig:vis-c2i-supp}
\end{figure*}

\begin{figure*}[!htbp]
    \centering
    \includegraphics[width=\linewidth]{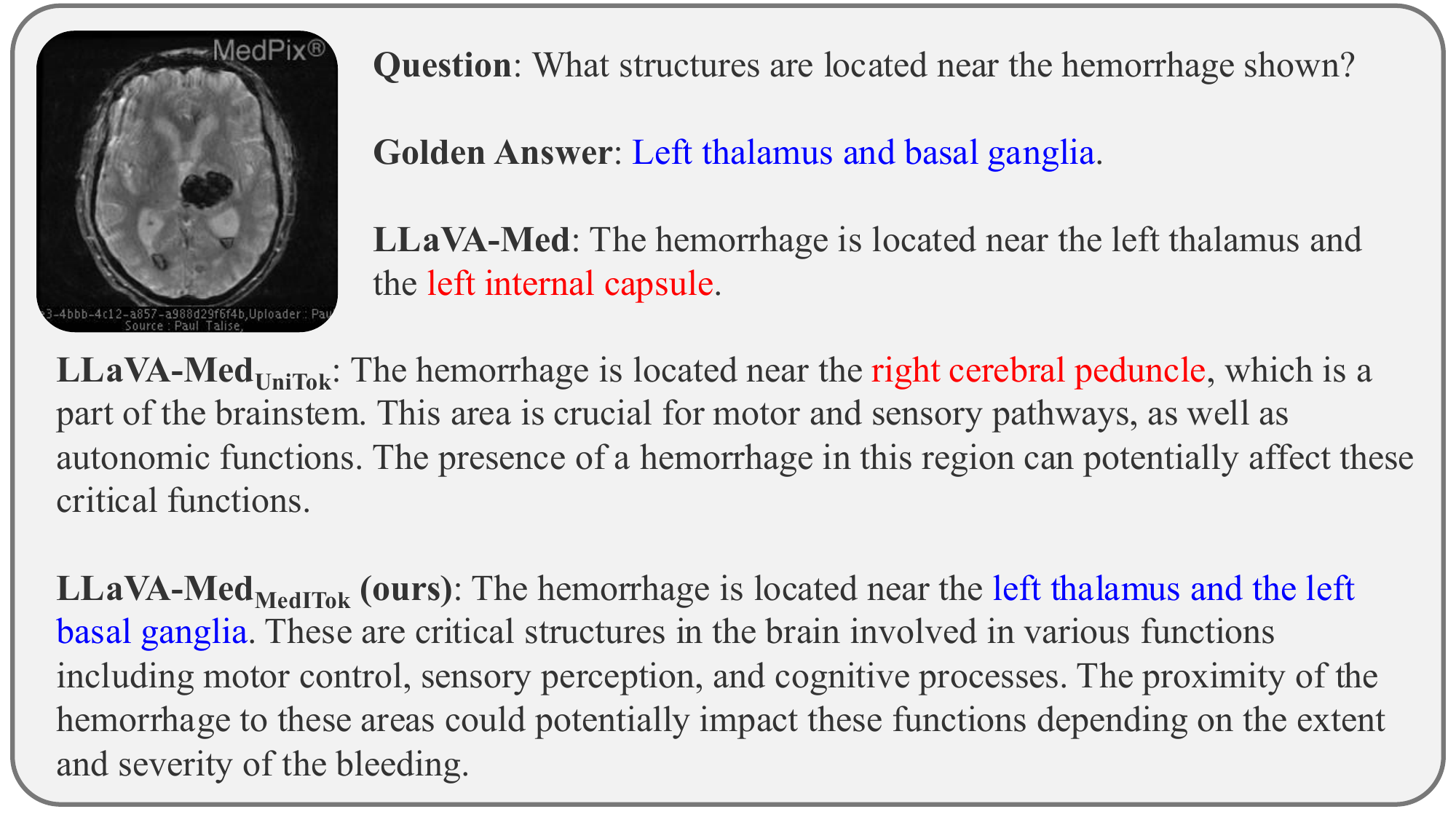}
    \caption{Visual question answering results on an MRI image.}
    \label{fig:vis-vqa1}
\end{figure*}

\begin{figure*}[!htbp]
    \centering
    \includegraphics[width=\linewidth]{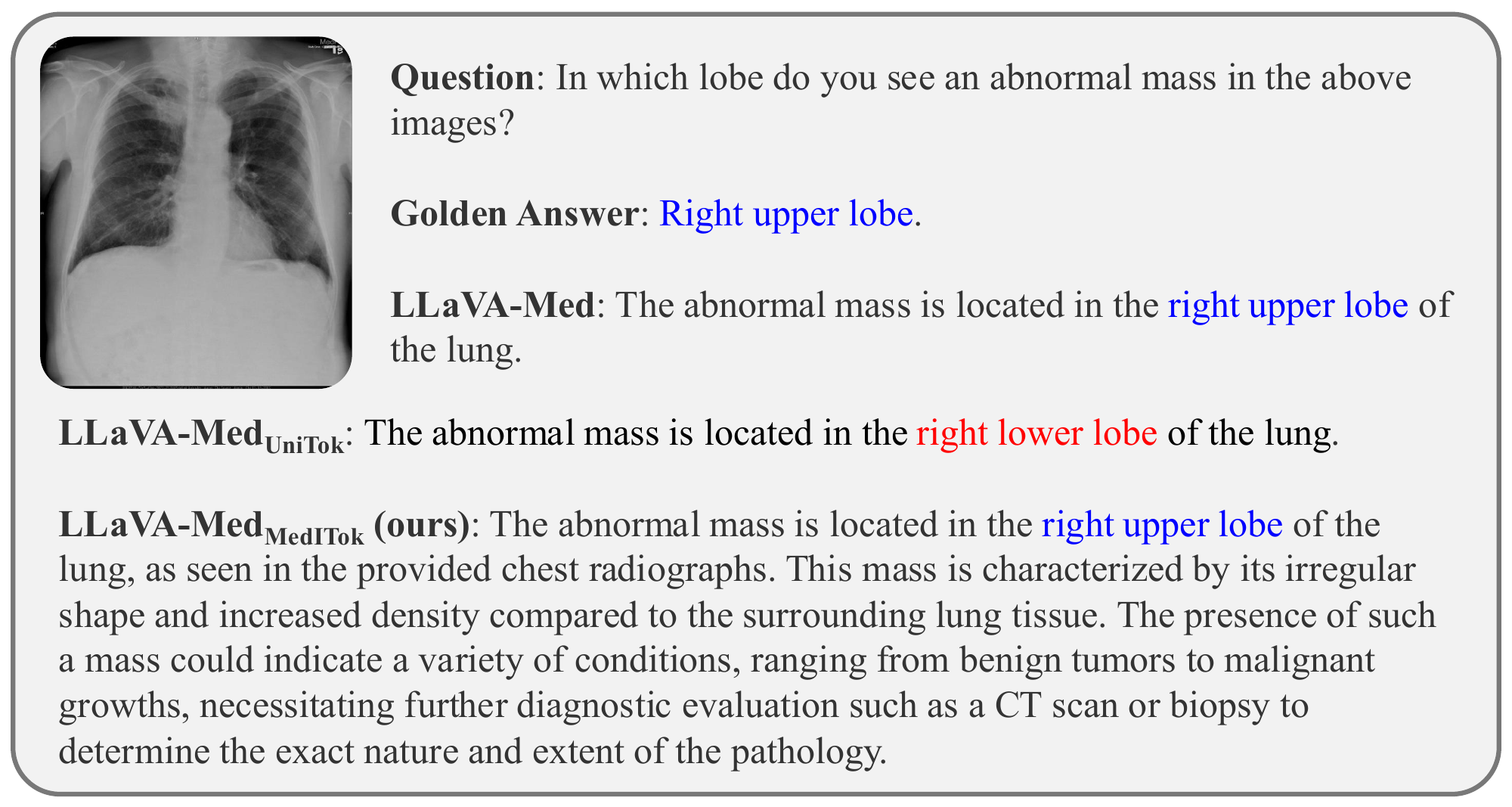}
    \caption{Visual question answering results on a chest X-ray image.}
    \label{fig:vis-vqa2}
\end{figure*}

\begin{figure*}[!htbp]
    \centering
    \includegraphics[width=\linewidth]{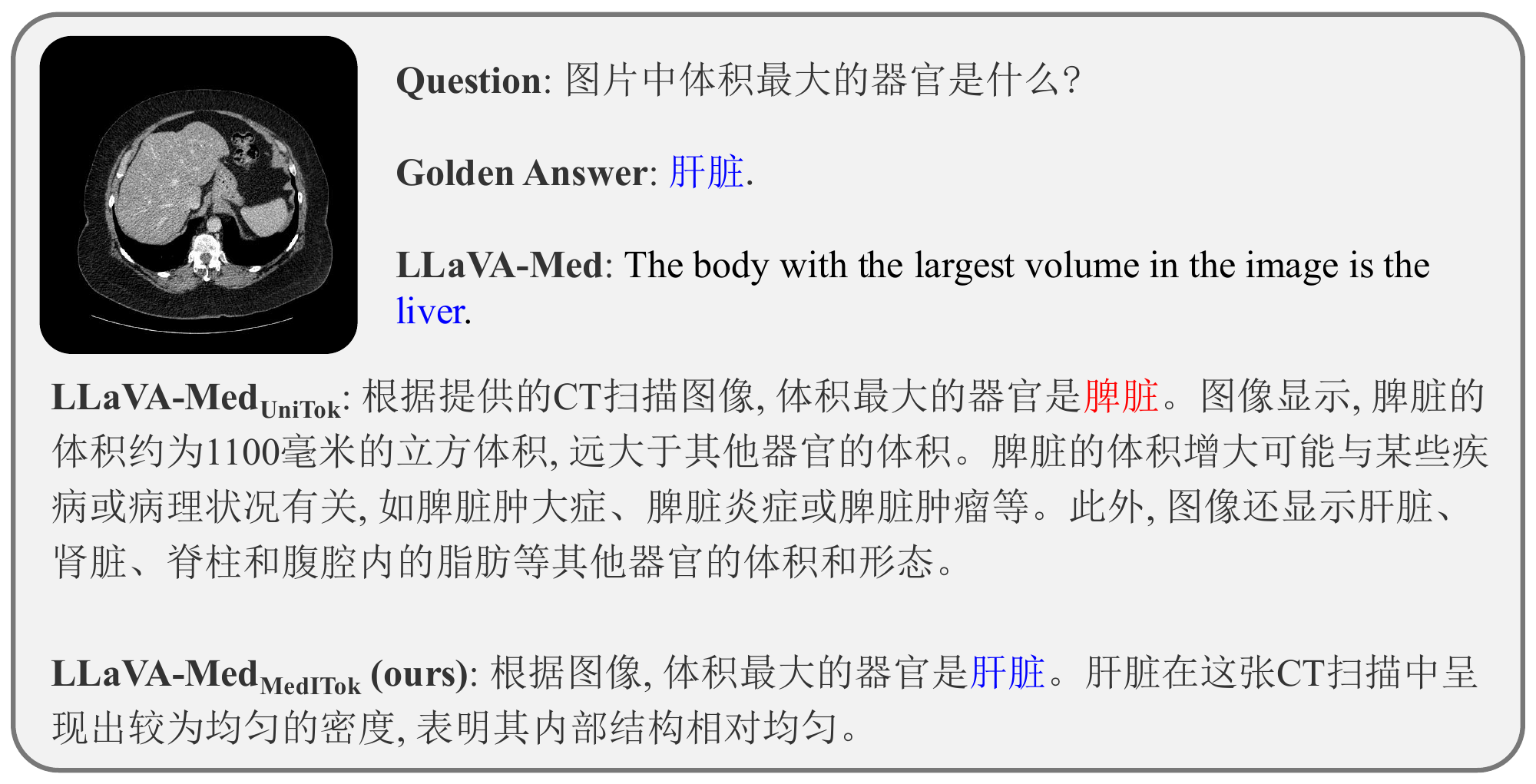}\caption{Visual question answering results on a CT image.}
    \label{fig:vis-vqa3}
\end{figure*}

Fig.~\ref{fig:vis-rec-supp} shows more examples for qualitative evaluation of medical image reconstruction, where \methodname achieves the best visual quality with lowest errors. Fig.~\ref{fig:vis-c2i-supp} compares the modality-conditioned synthesized images produced by different LlamaGen models. Notably, the LlamaGen model that adopts our \methodname as the visual tokenizer yields diverse and realistic medical images. Figs.~\ref{fig:vis-vqa1}--\ref{fig:vis-vqa3} presents the visual question answering results of LLaVA models that incoporate different visual tokenizers as their respective image encoder.

\begin{blockquote}
\small\emph{Return to: }
\hyperref[toc:appendix]{\textbf{Title Page}}
\end{blockquote}

\section{Discussion}\label{sec:supp:discussion}
This section synthesizes our design choices, positioning, limitations, and societal implications. 
We first justify the choice of discrete codebooks (Appendix~\ref{sec:supp:discussion-codebook}). We then situate \methodname relative to representative related works, clarifying differences in objectives, training, and latent space spaces (Appendix~\ref{sec:supp:discussion-related}). Next, we discuss current limitations of \methodname, and outline concrete avenues for future work (Appendix~\ref{sec:supp:discussion-limitation}). Finally, we reflect on broader impact and responsible use (Appendix~\ref{sec:supp:discussion-impact}).

\subsection{Choice of Discrete Codebooks}\label{sec:supp:discussion-codebook}
Our choice of discrete tokenization is driven by the goal of building a unified latent space that can power AR models across both image synthesis and interpretation tasks in the medical domain.

To that end, discrete tokens offer the following key advantages:
\begin{itemize}
    \item \textbf{Leveraging advances in AR modeling.} Discrete tokenization allows the medical community to directly benefit from the broader ecosystem of discrete-sequence modeling, \eg, unified training objectives, any-to-any modality transfer~\cite{zhan2024anygpt,chen2025janus}, and efficient decoding and infrastructure, which are not easily transferable to continuous tokenizers.

    \item \textbf{Unified latent space for visual synthesis and interpretation.} Discrete tokens act as a shared representational ``language'' across modalities. They support AR models that can both synthesize medical images and interpret them using a single AR backbone~\cite{lin2025healthgpt}. In contrast, continuous representations (\eg, VAEs, CLIP) typically lack this versatility, either being hard to decode (CLIP) or poorly aligned with semantic embeddings (VAE).

    \item \textbf{Seamless integration with different modalities.} Discrete visual tokens are natively compatible with discrete textual tokens, enabling direct multimodal fusion in AR models without additional heads or diffusion modules. This compatibility is critical for scaling medical AR models in the style of GPT-4o, where all modalities are treated as token sequences.
\end{itemize}

However, we also note the unique advantages of continuous visual tokenization, \eg, it reduces information loss and enables smooth latent space interpolation and morphing. Exploring a continuous unified latent space for medical images will be one of our future directions.

\subsection{Comparison with Related Works}\label{sec:supp:discussion-related}

We situate \methodname alongside the following related works: MedVAE~\cite{varma2025medvae}, VA-VAE~\cite{yao2025reconstruction}, and CLIP-based medical image tokenizers~\cite{zhang2022contrastive,wang2022medclip,huang2021gloria,huang2024enhancing,zhou2023transformer,zhou2022generalized}.

\subsubsection{MedVAE}
MedVAE is an effective continuous variational autoencoder (VAE) designed for efficient medical image interpretation. Our primary departure from MedVAE lies in where and how semantics are bound to the latent space. 
Before detailing the differences, we briefly describe the training stage of interest for MedVAE and \methodname: 
\begin{itemize}
    \item MedVAE first trains a continuous VAE, then freezes the VAE encoder and decoder and learns a lightweight projector whose output is optimized so that the BiomedCLIP image embedding of the projected latent matches the embedding of the input image via an $\ell_2$ loss, \ie, $\ell_2(\mathcal{E}_\mathrm{vis}(f(\mat{z})), \mathcal{E}_\mathrm{vis}(\mat{x}))$, where $\mathcal{E}_\mathrm{vis}$ denotes the pretrained BiomedCLIP vision encoder, $f$ is the projector, $\mat{x}$ is the input image, and $\mat{z}$ is the corresponding latent.
    
    \item \methodname utilizes $\mathcal{L}_\mathrm{contrastive}(f(\mat{z}), \mathcal{E}_\mathrm{txt}(\mat{t}))$ (or $\mathcal{L}_\mathrm{contrastive}(f(\mat{z}), \mathcal{E}_\mathrm{vis}(\mat{x}))$, as in the first stage), where $\mathcal{L}_\mathrm{contrastive}$ is the contrastive loss, and $\mat{t}$ denotes the caption. In either stage, the encoder and decoder of \methodname are trainable.
\end{itemize}

This clearly shows the following main differences:
\begin{enumerate}
    \item MedVAE enforces the latent $\mat{z}$ to be \emph{perceptually close} to the input image $\mat{x}$ under BiomedCLIP, which focuses more on improving the reconstruction fidelity, while \methodname aligns $\mat{z}$ to the embedding space of BiomedCLIP so the \methodname \emph{encodes more clinical semantics}.

    \item MedVAE keeps the VAE encoder and decoder frozen in the second stage, which can be viewed as treating semantics as post-hoc \emph{extraction} from a fixed latent. In contrast, \methodname \emph{injects} semantics into a discrete token space since the encoder and decoder of the tokenizer is \emph{both trainable}. 

    \item Since MedVAE focuses more on preserving structural details in radiological images, it did not utilize caption data for training and did not provide unified latent space for a wide range of downstream modalities and tasks. In contrast, by aligning latent tokens to BiomedCLIP embedding space, \methodname provides richer, fine-grained clinical semantics, which can be reflected in 
    Table~\ref{tab:main-cls}, 
    where \methodname shows significantly better performance than MedVAE on image classification tasks.
\end{enumerate}

\subsubsection{VA-VAE}
Another related work is VA-VAE~\cite{yao2025reconstruction}, which targets the reconstruction-generation trade-off in continuous VAE tokenizers for natural-image latent diffusion, proposing a single-stage joint reconstruction and alignment objective that aligns latents to a frozen vision foundation model to improve generative quality and training efficiency. The differences are as follows:
\begin{enumerate}
    \item Primary task. VA-VAE focuses on improved visual generation using semantic constraint in latent diffusion, whereas our work targets unified generation and interpretation (\eg, classification/VQA) across diverse medical modalities. This dual-use requirement drives our design choices.
    \item Methodology design. VA-VAE employs a single-stage objective to refine the latent space for better visual generation. In contrast, we use a two-stage curriculum to reach the unified goal while exploiting abundant unlabeled medical data. Moreover, VA-VAE uses cosine similarity as the alignment objective. However, as shown in the last chunk of Table~\ref{tab:abl-supp}, when directly adopted in our framework, such objective significantly degrades the medical image classification performance.
    \item Architecture. VA-VAE operates in a continuous VAE/diffusion setting; \methodname produces discrete, AR-ready tokens. Architecture is not the crux here, but this helps explain downstream usage differences.
    \item Community. VA-VAE contributes greatly to the field of general visual generation at designing effective VAEs. Our goal, however, is to democratize a foundation visual tokenizer for medical images to serve downstream applications, with effectiveness, scalability, and general usability for the medical image community.
\end{enumerate}

\subsubsection{CLIP-based}
\methodname also differs drastically from CLIP-based methods~\cite{zhang2022contrastive,wang2022medclip,huang2021gloria,huang2024enhancing,zhou2023transformer,zhou2022generalized}. 
The key distinction lies in the fundamental output and purpose of the models: CLIP-based works focus on representation learning for discrimination, whereas \methodname is designed for unified encoding of low-level structures and high-level semantics. Existing frameworks like ConVIRT~\cite{zhang2022contrastive}, GLoRIA~\cite{huang2021gloria}, MedCLIP~\cite{wang2022medclip}, and REFERS~\cite{zhou2022generalized} primarily utilize contrastive or cross-supervised objectives to align medical images with text into continuous vector embeddings. While effective for discriminative tasks like image classification, these models did not present reconstruction capability vital for low-level structure preserving and image synthesis. Similarly, other methods~\cite{huang2024enhancing,zhou2023transformer} focus on refining representations for accurate clinical diagnosis or granular feature matching, but they are not designed to reconstruct pixels or synthesize new images. In contrast, \methodname supports both high-fidelity image synthesis and interpretation within a single unified framework, a capability absent in the purely representation-focused baselines.

\subsection{Limitation and Future Directions}\label{sec:supp:discussion-limitation}

While \methodname demonstrates strong performance across multiple medical vision tasks, there remain important considerations and limitations that motivate future work.

\emph{First}, our two-stage training framework effectively balances structural fidelity and semantic alignment. However, optimizing simultaneously for both properties remains non-trivial. It is interesting and valuable to explore disentangling structural and semantic objectives during training~\cite{qu2024tokenflow} or jointly optimizing the tokenizer with a downstream model that unifies visual generation and interpretation~\cite{wang2025end}. We opt for the current two-stage design for its simplicity and effectiveness. 

\emph{Second}, although the current version of \methodname is designed mainly for 2D medical images across multiple imaging modalities, we have also shown that \methodname can be easily adapted to 3D medical tasks that require volume processing (Table~\ref{tab:ext-3d}). Nonetheless, \methodname could benefit from future advancement such as 3D native training or mixed training using 2D images and 3D volumes, as well as evaluation on more sophisticated tasks. 

\emph{Third}, due to resource constraints, our current experiments utilize 2.4 million image-caption pairs -- modest in scale compared to billion-scale training regimes in the general domain~\cite{ma2025unitok}. We believe that the proposed framework is scalable and can benefit significantly from larger and more diverse image-text corpora. Future efforts may explore combining public data with institution-curated pairs. 

In summary, while \methodname sets a strong foundation for unified medical visual tokenization, ongoing work is needed to address the above limitations. We envision that \methodname's flexible and expressive design can be extended to diverse downstream tasks. More broadly, we hope this work paves the way toward building scalable, general-purpose generative models that can advance medical AI and ultimately contribute to improving human health.

\subsection{Broader Impact}\label{sec:supp:discussion-impact}
This work presents a unified visual tokenizer tailored for medical images, offering a flexible and generalizable foundation for a wide range of medical AI applications. \methodname has the potential to accelerate the development of general-purpose medical AI systems and reduce task-specific engineering efforts. Its modular and pretrained nature also lowers the barrier for medical researchers to develop high-performance models with limited data and compute.

However, this progress also raises societal considerations. Insufficient training data may lead to biased models that underperform in underrepresented populations or clinical contexts. Additionally, the deployment of powerful downstream generative models in medicine, based on our \methodname, must be guided by strict ethical oversight to prevent misuse, misinformation, or over-reliance without clinical validation. We advocate for responsible development and interdisciplinary collaboration to ensure that such technologies benefit patients and healthcare systems.




\clearpage
\newpage

\newpage

\begin{table*}[htbp]
\centering
\caption{Medical Image Datasets in Stage 1 (Part 1 of 4).}
\label{tab:datasets-s1-part1}
\small
\begin{tabular*}{\textwidth}{p{0.39\textwidth}r|p{0.4\textwidth}r}
\toprule
\textbf{Dataset Name} & \textbf{Count} & \textbf{Dataset Name} & \textbf{Count} \\ 
\midrule
Rsna-Str-Pulmonary-Embolism-Detection~\cite{s1-rsna-str-pulmonary} & 5,604,627 & Bcnb-Task5~\cite{s1-bcnb} & 76,559 \\
Endovis2023-Surgtoolloc~\cite{s1-endovis2023-surgtoolloc} & 3,710,685 & Bcnb-Task1-0~\cite{s1-bcnb} & 76,558 \\
Panda~\cite{s1-panda} & 1,616,913 & Bcnb-Task6~\cite{s1-bcnb} & 76,558 \\
Mela~\cite{s1-mela-p1,s1-mela-p2,s1-mela-p3,s1-mela-p4} & 1,403,843 & Msd-Liver~\cite{s1-msd} & 76,395 \\
Ixi~\cite{s1-ixi} & 924,870 & Ct-Org~\cite{s1-ct-org} & 76,195 \\
Ribfrac2020~\cite{s1-ribfrac2020} & 810,265 & Endovis-2021-Petraw~\cite{s1-endovis-2021-petraw} & 75,718 \\
Radimagenet~\cite{s1-radimagenet} & 779,768 & Head-Neck-Pet-Ct~\cite{s1-head-neck-pet-ct} & 75,109 \\
Autopet~\cite{s1-autopet} & 590,785 & Ctspine1K~\cite{s1-ctspine1k} & 72,835 \\
Brats2023-Gli~\cite{s1-brats2023-gli} & 513,263 & Bcnb-Task1-2~\cite{s1-bcnb} & 71,991 \\
Atm2022~\cite{s1-atm2022} & 501,147 & Lndb~\cite{s1-lndb} & 70,292 \\
Lidc-Idri-All-Ct~\cite{s1-lidc-idri-all-ct} & 474,076 & Cptac-Hnscc~\cite{s1-cptac-hnscc} & 69,731 \\
Luna16~\cite{s1-luna16} & 431,694 & Lung-Pet-Ct-Dx~\cite{s1-lung-pet-ct-dx} & 66,564 \\
Brats2023-Men~\cite{s1-brats2023-men} & 384,425 & Anti-Pd-1-Melanoma~\cite{s1-anti-pd-1-melanoma} & 65,411 \\
Mimic-Cxr~\cite{johnson2019mimic} & 377,110 & Nsclc-Cetuximab~\cite{s1-nsclc-cetuximab} & 64,730 \\
Qin-Headneck~\cite{s1-qin-headneck} & 307,946 & Anode09~\cite{s1-anode09} & 63,250 \\
Biomedica~\cite{lozano2025biomedica} & 291,155 & Opc-Radiomics & 62,726 \\
Flare22~\cite{s1-flare22} & 280,531 & Acrin-Nsclc-Fdg-Pet~\cite{s1-acrin-nsclc-fdg-pet} & 62,701 \\
Braintumour~\cite{s1-braintumour} & 263,310 & Sln-Breast~\cite{s1-sln-breast} & 61,968 \\
Chexpertplus~\cite{s1-chexpertplus} & 223,460 & Bcnb-Task2~\cite{s1-bcnb} & 61,828 \\
Totalsegmentator-Dataset~\cite{s1-totalsegmentator} & 218,477 & Msd-Lung~\cite{s1-msd} & 61,117 \\
Pediatric-Ct-Seg~\cite{s1-pediatric-ct-seg,s1-pediatric-ct-seg-p2} & 204,602 & Bcnb-Task1-3~\cite{s1-bcnb} & 59,521 \\
Acrin6668~\cite{s1-acrin6668} & 188,098 & Midrc-Ricord-1B~\cite{s1-midrc-ricord} & 59,247 \\
Covid-19-Ny-Sbu~\cite{s1-covid-19-ny-sbu} & 185,668 & Bcnb-Task1-4~\cite{s1-bcnb} & 59,091 \\
Bracs~\cite{s1-bracs} & 177,712 & Learn2Reg2022-L2R-Task1-Oasis~\cite{s1-learn2reg} & 57,984 \\
Abdomenct1K~\cite{s1-abdomenct1k} & 172,963 & Amos2022~\cite{s1-amos} & 56,217 \\
Bone-Marrow-Cytomorphology~\cite{s1-bone-marrow-cytomorphology} & 171,378 & Learn2Reg22-L2R-Oasis~\cite{s1-learn2reg} & 52,992 \\
Ctpelvic1K~\cite{s1-ctpelvic1k} & 127,315 & Cataract101~\cite{s1-cataract101} & 52,676 \\
Parse22~\cite{s1-parse22} & 122,629 & Brats2023-Ped~\cite{s1-brats2023-ped} & 51,769 \\
Nih-Chest-X-Rays~\cite{s1-nih-chest-x-rays} & 112,115 & Vestibular-Schwannoma-Seg~\cite{s1-vestibular-schwannoma-seg} & 51,575 \\
Lits~\cite{s1-lits} & 107,056 & Midrc-Ricord-1A~\cite{s1-midrc-ricord} & 50,913 \\
Hnscc~\cite{s1-hnscc,s1-hnscc-p2} & 101,861 & Lc25000~\cite{s1-lc25000} & 50,000 \\
Airogs~\cite{s1-airogs} & 101,280 & Cptac-Luad~\cite{s1-cptac-luad} & 48,952 \\
Head-Neck-Cetuximab~\cite{s1-head-neck-cetuximab} & 100,356 & Ct-Covid-19-August2020~\cite{s1-ct-covid-19-august2020} & 48,791 \\
Brats2023-Met~\cite{s1-brats2023-met} & 93,775 & Fastpet-Ld~\cite{s1-fastpet-ld} & 48,097 \\
Acrin-Flt-Breast~\cite{s1-acrin-flt-breast} & 91,948 & Oasis2~\cite{s1-oasis2} & 47,744 \\
Bcnb-Task4~\cite{s1-bcnb} & 89,894 & Osic-Pul-Fib-Pro~\cite{s1-osic-pul-fib-pro} & 46,014 \\
Covidx-Cxr-4~\cite{s1-covidx-cxr-4} & 84,802 & Anti-Pd-1-Lung~\cite{s1-anti-pd-1-lung} & 45,497 \\
Nlst~\cite{s1-nlst} & 79,194 & Tcga-Luad~\cite{s1-tcga-luad} & 45,049 \\
Cad-Pe~\cite{s1-cad-pe} & 78,583 & Isic2020~\cite{s1-isic2020} & 44,106 \\
Bcnb-Task3~\cite{s1-bcnb} & 76,559 & Longitudinal-multiple-sclerosis-lesion-segmentation~\cite{s1-longitudinal} & 41,984 \\

\bottomrule
\end{tabular*}
\end{table*}

\begin{table*}[htbp]
\centering
\caption{Medical Image Datasets in Stage 1 (Part 2 of 4).}
\label{tab:datasets-s1-part2}
\small
\begin{tabular*}{\textwidth}{p{0.39\textwidth}r|p{0.4\textwidth}r}
\toprule
\textbf{Dataset Name} & \textbf{Count} & \textbf{Dataset Name} & \textbf{Count} \\ 
\midrule
Covid-19-Ar~\cite{desai2020chest} & 41,664 & Lysto~\cite{10296868} & 19,990 \\
Glis-Rt~\cite{https://doi.org/10.7937/tcia.t905-zq20} & 41,143 & Cas2023~\cite{cas2023} & 19,200 \\
Mura~\cite{s1-mura} & 39,939  & Tcga-Ov~\cite{https://doi.org/10.7937/k9/tcia.2016.ndo1mdfq} & 19,077 \\
Spie-Aapm~\cite{https://doi.org/10.7937/k9/tcia.2015.uzlsu3fl} & 39,670 & Sicapv2~\cite{silva2020going} & 18,783 \\
Tcga-Lusc~\cite{https://doi.org/10.7937/k9/tcia.2016.tygkkfmq} & 38,998 & Vin-Big-Data~\cite{vinbigdata-chest-xray-abnormalities-detection} & 17,999 \\
Atlas-2~\cite{s1-atlas-2} & 38,400 & Wmh~\cite{8669968} & 16,896 \\
Spie-Aapm-Lung-Ct-Challenge~\cite{https://doi.org/10.7937/k9/tcia.2015.uzlsu3fl} & 38,373 & Fizpatrick17K~\cite{groh2021evaluating,groh2022towards} & 16,577 \\
M2Cai16-Tool~\cite{jin2018tool} & 37,314 & Chest-Image-Pneum~\cite{siim-acr-pneumothorax-segmentation} & 15,251 \\
Hyperkvasir~\cite{Borgli2020} & 36,329 & C-Nmc-2019~\cite{https://doi.org/10.7937/tcia.2019.dc64i46r} & 15,105 \\
Brats-Tcga-Gbm~\cite{https://doi.org/10.7937/k9/tcia.2017.klxwjj1q} & 35,770 & Covid-19-20~\cite{roth2022rapid} & 15,045 \\
Lld-Mmri2023~\cite{LLD-MMRI2023} & 35,751 & Aod-14800~\cite{Aod-14800} & 14,805 \\
Diabetic~\cite{Tianchi_Diabetic} & 35,059 & Aapm-Rt-Mac~\cite{https://doi.org/10.7937/tcia.2019.bcfjqfqb} & 14,080 \\
Eyepacs~\cite{diabetic-retinopathy-detection} & 35,059 & Mindboggle~\cite{10.1371/journal.pcbi.1005350} & 12,575 \\
Ranzcr-Clip~\cite{ranzcr-clip-catheter-line-classification} & 33,664 & Siim-Acr-Pneumothorax~\cite{siim-acr-pneumothorax-segmentation} & 12,053 \\
Isic2019~\cite{codella2018skinlesionanalysismelanoma} & 33,541 & Chest-X-Ray-Images-With-Pneumothorax-Masks~\cite{siim-acr-pneumothorax-segmentation} & 12,047 \\
Verse20~\cite{SEKUBOYINA2021102166} & 32,944 & Han-Seg~\cite{https://doi.org/10.1002/mp.16197} & 11,939 \\
Covidxcxr-2~\cite{Wang2020} & 31,238 & Valdo-Task1~\cite{s1-valdo} & 11,915 \\
Lola11~\cite{van_ginneken_2021_4708800} & 30,207 & Valdo-Task3~\cite{s1-valdo} & 11,915 \\
Rsna-Pdc~\cite{rsna-pneumonia-detection-challenge} & 29,684 & Cptac-Ucec~\cite{https://doi.org/10.7937/k9/tcia.2018.3r3juisw} & 11,595 \\
C4Kc-Kits~\cite{https://doi.org/10.7937/tcia.2019.ix49e8nx} & 28,843 & Tcga-Stad~\cite{https://doi.org/10.7937/k9/tcia.2016.gdhl9kim} & 11,204 \\
Word~\cite{s1-word} & 27,154 & Ultrasound-Nerve-Segmentation~\cite{ultrasound-nerve-segmentation} & 11,143 \\
Acrin-Hnscc-Fdg-Pet-Ct~\cite{https://doi.org/10.7937/k9/tcia.2016.jqejzzng} & 27,117  & Msseg08~\cite{styner20083d} & 10,965 \\
Kits2021~\cite{heller2020state} & 26,503 & Wsss4Luad~\cite{han2022wsss4luad} & 10,091 \\
Exact09~\cite{s1-exact09} & 25,560 & Medfm-Colon-2023~\cite{wang2023real} & 10,009 \\
Bcnb-Task1-1~\cite{s1-bcnb} & 25,370 & Knee-Osteoarthritis-Dataset~\cite{Knee-Osteoarthritis-Dataset} & 9,766 \\
Surgvisdom~\cite{s1-surgvisdom} & 24,360 & Segthor~\cite{lambert2020segthor} & 9,661 \\
Brats-Tcga-Lgg~\cite{https://doi.org/10.7937/k9/tcia.2017.gjq7r0ef} & 23,336 & Brain-Ptm~\cite{avital2019neural,nelkenbaum2020automatic} & 9,600 \\
Tcga-Ucec~\cite{https://doi.org/10.7937/k9/tcia.2016.gkj0zwac} & 22,946 & Msd-Colon~\cite{s1-msd} & 9,191 \\
Tcga-Kirc~\cite{https://doi.org/10.7937/k9/tcia.2016.v6pbvtdr} & 22,644 & Covid19Ctscans~\cite{ma_jun_2020_3757476} & 9,119 \\
Cptac-Sar~\cite{https://doi.org/10.7937/tcia.2019.9bt23r95} & 22,432 & Cholect50~\cite{Nwoye_2023} & 8,919 \\
Crossmoda2023~\cite{Dorent_2023} & 21,981 & Msd-Pancreas~\cite{s1-msd} & 8,666 \\
Cptac-Cm~\cite{https://doi.org/10.7937/k9/tcia.2018.odu24gze} & 21,867 & Fumpe~\cite{masoudi2018new} & 8,402 \\
Brats2023-Ssa~\cite{adewole2023braintumorsegmentationbrats} & 20,910 & Lctsc~\cite{https://doi.org/10.7937/k9/tcia.2017.3r3fvz08} & 8,300 \\
Pancreas-Ct~\cite{https://doi.org/10.7937/k9/tcia.2016.tnb1kqbu} & 20,709 & Ct-Vs-Pet-Ventilation-Imaging~\cite{https://doi.org/10.7937/3ppx-7s22} & 8,252 \\
Vessel2012~\cite{RUDYANTO20141217} & 20,442 & Head-Neck-Radiomics-Hn1~\cite{https://doi.org/10.7937/tcia.2019.8kap372n} & 8,161 \\
Yangxi~\cite{chi_liu_2019_3393265} & 20,394 & Qin-Breast~\cite{https://doi.org/10.7937/k9/tcia.2016.21juebh0} & 8,051 \\
Msseg2016~\cite{commowick2018objective} & 20,352 & Chaos-Task-4~\cite{CHAOS2021} & 7,977 \\
Oia-Odir~\cite{ioa-odir} & 19,992 & Pannuke~\cite{gamper2019pannuke,gamper2020pannuke} & 7,810 \\
\bottomrule
\end{tabular*}
\end{table*}

\begin{table*}[htbp]
\centering
\caption{Medical Image Datasets in Stage 1 (Part 3 of 4).}
\label{tab:datasets-s1-part3}
\small
\begin{tabular*}{\textwidth}{p{0.39\textwidth}r|p{0.4\textwidth}r}
\toprule
\textbf{Dataset Name} & \textbf{Count} & \textbf{Dataset Name} & \textbf{Count} \\ 
\midrule
Sppin2023~\cite{buser2025automated} & 7,616 & Pad-Ufes-20~\cite{pacheco2020pad} & 2,298 \\
Atlas2023~\cite{data8050079} & 7,364 & Msd-Spleen~\cite{s1-msd} & 2,169 \\
Msd-Hepaticvessel~\cite{s1-msd} & 6,859 & Breakhis-100X~\cite{spanhol2015dataset} & 2,081 \\
Mmwhs~\cite{zhuang2018multivariate} & 6,400 & Breakhis-200X~\cite{spanhol2015dataset} & 2,011 \\
Hsa-Nrl~\cite{zhu2021hard} & 6,160 & Breakhis-40X~\cite{spanhol2015dataset}& 1,991 \\
Coronahack~\cite{coronahack2020} & 5,933 & Breakhis-400X ~\cite{spanhol2015dataset}& 1,820 \\
Rus-Chn~\cite{s1-rus-chn} & 5,921 & Cptac-Pda~\cite{national2018clinical} & 1,792 \\
Dhrf~\cite{derbi2022fundus} & 5,680 & Tiger-Wsirois-Roi-Level-Tissue-Cells~\cite{vanrijthoven2022tiger} & 1,775 \\
Aptos2019-Blindness-Detection~\cite{aptos2019}& 5,590 & Breast-Diagnosis~\cite{wolberg1995wdbc} & 1,656 \\
Curious2019~\cite{xiao2019evaluation}& 5,376 & Cmb-Gec~\cite{cmb_gec_2022} & 1,625 \\
Cmb-Mel~\cite{cmb_mel_2022} & 5,289 & Riga-Dataset~\cite{RIGA} & 1,617 \\
Clust15-2D~\cite{DeLuca2018Evaluation}& 5,255 & Refuge2-Cls~\cite{fang2022refuge2} & 1,600 \\
Cmmd~\cite{cui2021cmmd} & 5,202 & Harvardglaucoma-1547~\cite{kim2018machine} & 1,544 \\
Tcga-Hnsc~\cite{Zuley2016TCGAHNSC} & 5,172 & Tcga-Kich~\cite{Linehan2016TCGAKICH} & 1,484 \\
Continuous-Registration-Task3~\cite{baheti2021brain} & 5,120 & Papilledema~\cite{papilledema_dataset_2020} & 1,369 \\
Messeg~\cite{commowick2018objective}& 5,120 & Continuous-Registration-Task6~\cite{hering2022learn2reg} & 1,280 \\
Node21~\cite{sogancioglu2024nodule}& 4,882 & Isbi2016-Part3~\cite{gutman2016skin} & 1,279 \\
Conic2022~\cite{Graham2022CoNIC}& 4,870 & Isic2016-Task1~\cite{gutman2016skin} & 1,279 \\
Lag-4854 ~\cite{li2019large}& 4,854 & Fusc2021~\cite{wang2024fuseg} & 1,210 \\
Medfm-Chestdr-2023~\cite{MedFM2023ChestDR} & 4,848 & Hvsmr-2016~\cite{pace2015interactive} & 1,152 \\
Stageii-Colorectal-Ct~\cite{tong2022abdominal}& 4,672 & Osteosarcoma-Tumor-Assessment~\cite{leavey2019osteosarcoma} & 1,143 \\
Naf-Prostate~\cite{Kurdziel2015NaF}& 4,664 &  Isic2016-Task2B-Globules~\cite{gutman2016skin} & 1,142 \\
Chest-X-Ray-Pa~\cite{asraf2021covid19}& 4,574 & Isic2016-Task2B-Streaks~\cite{gutman2016skin} & 1,142 \\
Lungct-Diagnosis~\cite{grove2015lungct} & 4,155 &  Jsiec~\cite{cen2021automatic}& 997 \\
Covid19-Radio-Data~\cite{chowdhury2020can}& 3,886 & Isles2022~\cite{hernandez2022isles}& 938 \\
Structseg2019-Subtask1~\cite{StructSeg2019Task} & 3,634 & Covid-19-Ct-Cxr-Det~\cite{peng2020covid} & 929 \\
Structseg2019-Subtask4~\cite{StructSeg2019Task} & 3,634 & Covid-19-Ct-Cxr~\cite{peng2020covid}& 918 \\
Structseg2019-Subtask2~\cite{StructSeg2019Task}  & 3,413 & E-Ophta~\cite{decenciere2013teleophta} & 905 \\
Qin-Lung-Ct~\cite{kalpathy2015qinlungct} & 3,586 & Dao-Slocpasa~\cite{chiu2013automatic}& 840 \\
Structseg2019-Subtask3~\cite{StructSeg2019Task}& 3,413 & Continuous-Registration-Task5~\cite{KLEIN2009786} & 813 \\
Tcga-Coad~\cite{cancer2012comprehensive}& 3,093 & Fives~\cite{Jin2022} & 800 \\
Tcga-Prad~\cite{abeshouse2015molecular}& 3,007 & Segpc2021~\cite{7np1-2q42-21} & 773 \\
Bidr-2838~\cite{islam2021deep}& 2,838 & Paraguay-757~\cite{veronica_elisa_castillo_benitez_2021_4891308} & 757 \\
Refuge2~\cite{fang2022refuge2}& 2,800 & Mudi2019~\cite{10.1007/978-3-030-52893-5_17} & 695 \\
Cptac-Ccrcc~\cite{https://doi.org/10.7937/k9/tcia.2018.oblamn27} & 2,798 & Pulmonary-Chest-X-Ray-China~\cite{6616679,6663723} & 662 \\
Isic2017~\cite{DBLP:journals/corr/abs-1710-05006} & 2,748 & Glaucoma-Detection~\cite{GlaucomaDetection} & 650 \\
Verse19~\cite{sekuboyina2021verse} & 2,650 & Beh-634~\cite{9664528} & 634 \\
Palm19~\cite{fang2024open} & 2,379 &  &  \\

\bottomrule
\end{tabular*}
\end{table*}

\begin{table*}[htbp]
\centering
\caption{Medical Image Datasets in Stage 1 (Part 4 of 4).}
\label{tab:datasets-s1-part4}
\small
\begin{tabular*}{\textwidth}{p{0.39\textwidth}r|p{0.4\textwidth}r}
\toprule
\textbf{Dataset Name} & \textbf{Count} & \textbf{Dataset Name} & \textbf{Count} \\ 
\midrule
Retina-Cataract-Dataset~\cite{RetinaDataset} & 601 & Orvs~\cite{sarhan2021transfer} & 202 \\
Idrid~\cite{2020idrid} & 597 & Gamma3~\cite{wu2023gamma} & 200 \\
Sz-Cxr~\cite{SzCxr} & 566 & Fund-179~\cite{s1-fund-179} & 179 \\
Cmb-Pca~\cite{CmbPca} & 532 & Drac2022-Taska2~\cite{qian2023drac} & 174 \\
Crass~\cite{Crass} & 518 & Drac2022-Taska3~\cite{qian2023drac} & 174 \\
Herlev~\cite{jantzen2005pap} & 504 & Tcga-Read~\cite{tcgaread} & 168 \\
Papila~\cite{kovalyk2022papila} & 488 & Glas~\cite{glas} & 165 \\
Rimonedl~\cite{RIMONEDLImageAnalStereol2346} & 485 & Drac2022-Taska1~\cite{qian2023drac} & 151 \\
Fetoscopy-Placenta-Dataset~\cite{bano2020deep} & 482 & Tiger-Wsirois-Roi-Level-Tissue-Bcss~\cite{amgad2019structured} & 151 \\
Tcga-Blca~\cite{tcgaread} & 439 & Tcga-Lgg~\cite{tcgaread} & 145 \\ 
Drimdb~\cite{prentavsic2013diabetic} & 428 & Pulmonary-Chest-X-Ray-Montgomery~\cite{Montgomeryset,2014lungseg} & 138 \\
Toxofundus~\cite{cardozo2023dataset,alam2023benchmarking} & 411 & Bcss~\cite{amgad2019structured} & 121 \\ 
Adam~\cite{adam} & 400 & Drishti-Gs-Cup~\cite{Drishti} & 101 \\
Ph2~\cite{mendoncca2015ph2} & 400 & Drishti-Gs-Od~\cite{Drishti} & 101 \\
Crown~\cite{vos2024results} & 384 & Avn~\cite{s1-avn} & 90 \\
Rose~\cite{ma2021rose} & 348 & Jsrt-Lung~\cite{Jsrt} & 60 \\ 
Mias~\cite{pisano2005digital} & 322 & Breast-Cancer-Cell-Seg~\cite{BreastCancerCell} & 58 \\
Covid-19-Image-Dataset~\cite{sohan2020so} & 317 & Monuseg~\cite{Monuseg} & 51 \\
Gamma~\cite{wu2023gamma} & 300 & Hrf~\cite{budai2013robust} & 45 \\
Monusac20~\cite{Monusac20} & 283 & Drhagis~\cite{holm2017dr} & 40 \\
Rod~\cite{el2019raman} & 281 & Drive~\cite{drive} & 40 \\
Jsrt~\cite{Jsrt} & 247 & Rite~\cite{rite} & 40 \\
Jsrt-Gender-Cls~\cite{Jsrt} & 247 & Hrf-Quality-Cls~\cite{budai2013robust} & 36 \\
Tcga-Sarc~\cite{tcgaread} & 241 & Retinacheck~\cite{dashtbozorg2016retinacheck} & 30 \\
Crag~\cite{crag} & 213 & Olives-Fundus-Photography~\cite{prabhushankar2022olives} & 14 \\
Panda-Radboud~\cite{nir2018automatic} & 206 & Occmcpv~\cite{chen2024segmentation} & 8 \\
\bottomrule
\end{tabular*}
\end{table*}

\begin{table*}[htbp]
\centering
\caption{Medical Image Datasets in Stage 2.}
\label{tab:datasets-s2-part1}
\small
\begin{tabular*}{\textwidth}{p{0.39\textwidth}r|p{0.4\textwidth}r}
\toprule
\textbf{Dataset Name} & \textbf{Count} & \textbf{Dataset Name} & \textbf{Count} \\ 
\midrule
Biomedica~\cite{lozano2025biomedica} & 1,216,529 & Mimic-Cxr~\cite{johnson2019mimic} & 107,684 \\
Gmai-Vl-5.5M~\cite{li2024gmai} & 671,824 & Rocov2~\cite{ruckert2024rocov2} & 59,212 \\
Medicat~\cite{subramanian2020medicat} & 204,772 & Pmc-Oa~\cite{lin2023pmcoa} & 36,386 \\
Llava-Med-Instruct-Fig-Captions~\cite{li2023llavamed} & 122,843 & Mm-Retinal~\cite{wu2024mmretinal} & 3,577 \\
\bottomrule
\end{tabular*}
\end{table*}

\clearpage

\begin{table*}[htbp]
\centering
\caption{Medical Image Datasets for Image Reconstruction Evaluation.}
\label{tab:datasets-recon-part1}
\small
\begin{tabular*}{\textwidth}{p{0.4\textwidth}r|p{0.4\textwidth}r}
\toprule
\textbf{Dataset Name} & \textbf{Count} & \textbf{Dataset Name} & \textbf{Count} \\ 
\midrule
Ivygap-Radiomics~\cite{pati2020ivygapradiomics} & 8,456 & Monkeypox~\cite{ali2022monkeypox} & 802 \\
Chestx-Det~\cite{lian2021chestxdet} & 3,578 & Breast-Ultrasound-Images-Dataset~\cite{al2020busi} & 647 \\
Aapm-lowdose-ct~\cite{mccollough2017aapmldct} & 3,413 & Ddti~\cite{pedraza2015ddti} & 637 \\
Btcv-Cervix~\cite{landman2015btcv} & 3,039 & Hie2023~\cite{bao2025hie} & 554 \\
Surgt~\cite{cartucho2024surgt} & 2,933 & Digestpath19-Cls~\cite{da2022digestpath} & 455 \\
Silver07~\cite{heimann2009silver07} & 2,291 & EndoCV2020-EDD~\cite{ali2020endocv20edd} & 386 \\
Derm7Pt~\cite{kawahara2018derm7pt} & 2,013 & Mednode~\cite{giotis2015mednode} & 170 \\
Messidor~\cite{decenciere2014messidor} & 1,748 & Gleason~\cite{nir2018gleason} & 103 \\
Rsna-Bone-Age~\cite{halabi2019rsnabone} & 1,596 & Consep~\cite{graham2019consep} & 41 \\
Hmc-Qu~\cite{kiranyaz2020hmcqu} & 1,269 & Chase~\cite{fraz2012chasedb} & 28 \\
Covidgr~\cite{tabik2020covidgr} & 852 & Stare~\cite{hoover2000stare} & 20 \\
\bottomrule
\end{tabular*}
\end{table*}


\begin{table*}[htbp]
\centering
\caption{Downstream Medical Vision Tasks Datasets. ``CLS'': classification. ``SYN'': image synthesis. ``VQA'': visual question answering.}
\label{tab:datasets-downstream}
\begin{tabular}{lrrlll}
\toprule
\textbf{Dataset} & \textbf{Train} & \textbf{Test} & \textbf{Modality} & \textbf{Task Type} & \textbf{Classes} \\ 
\midrule
Kermany~\cite{kermany2018pneumnist} & 4,708 & 1,148 & X-ray & CLS & 2 \\
NCT-CRC~\cite{kather2019pathmnist} & 89,996 & 500 & pathology & CLS; SYN & 9 \\
NIH-ChestXray14~\cite{wang2017chestmnist} & 78,468 & 500 & X-ray & SYN & 14 \\
PBC~\cite{acevedo2020bloodmnist} & 11,959 & 500 & microscopy & SYN & 8 \\
ISIC2018~\cite{tschandl2018dermamnistham10000,codella2019dermamnistskin} & 7,007 & 500 & dermoscopy & CLS; SYN & 7 \\
DeepDRID~\cite{liu2022retinamnistdeepdrid} & 1,080 & 500 & fundus photography & CLS; SYN & 5 \\
BUSI~\cite{al2020busi} & 546 & 234 & ultrasound & CLS; SYN & 2 \\
Derm12345~\cite{yilmaz2024derm12345} & 9,860 & 2,485 & dermoscopy & SYN & 5 \\
Pubmed-Vision-Caption~\cite{chen2024huatuogpt} & 555,103 & 0 & Unknown & VQA & -- \\
Pubmed-Vision-VQA~\cite{chen2024huatuogpt} & 100,000 & 0 & Unknown & VQA & -- \\
VQARAD-Test~\cite{lau2018vqarad} & 0 & 451 & Unknown & VQA & -- \\
Slake-Test~\cite{liu2021slake} & 0 & 2,094 & Unknown & VQA & -- \\
Slake-Val~\cite{liu2021slake} & 0 & 2,099 & Unknown & VQA & -- \\
\bottomrule
\end{tabular}
\end{table*}

\clearpage
\newpage


\bibliographystyle{splncs04}
\bibliography{resources/bibs/method.bib, resources/bibs/dataset.bib}
\end{document}